\documentclass[prl,twocolumn,notitlepage,longbibliography]{revtex4-1}
\usepackage{amsmath}
\usepackage{amssymb}

\usepackage[unicode=true,colorlinks=true,citecolor=blue,urlcolor=blue]{hyperref}

\usepackage{bm}
\usepackage{color}
\usepackage{xcolor}
\usepackage{epsfig}

\usepackage[toc,page]{appendix}

\newcommand{\nocontentsline}[3]{}
\newcommand{\tocless}[2]{\bgroup\let\addcontentsline=\nocontentsline#1{#2}\egroup}
 
\usepackage[normalem]{ulem}

\renewcommand {\Im}{\mathop\mathrm{Im}\nolimits}
\renewcommand {\Re}{\mathop\mathrm{Re}\nolimits}
\newcommand {\rmi}{{\rm i}}
\newcommand {\rmd}{{\rm d}}
\newcommand {\sign}{\mathop{\mathrm{sign}}\nolimits}
\newcommand {\e}{{\rm e}}
\newcommand {\eps}{\varepsilon}

\newcommand {\rot}{\mathop\mathrm{rot}\nolimits}

\begin{document}
\title{Optomechanical tension and crumpling of resonant membranes}
\author{{I.D. Avdeev, A.N. Poddubny, and A.V. Poshakinskiy}}
\affiliation{Ioffe Institute, St. Petersburg 194021, Russia}
\email{poshakinskiy@mail.ioffe.ru}
\date{\today}
\begin{abstract}
We predict that illumination by a plane electromagnetic wave {of optically resonant membranes, such as graphene or monolayers of transition metal dichalcogenides,  directly affects their mechanical tension.}
The induced optomechanical tension is anisotropic and, depending on the spectral detuning  from the resonance, can be both positive and negative. In the latter case, it can overcome the bending rigidity of the membrane leading to transition to the crumpled phase. 
The instability caused by optomechanical heating of flexural vibrations is also considered. 
\end{abstract}
\maketitle 

\paragraph{Introduction.}  Membranes demonstrate rich mechanical phenomena, including  non-Hookean elasticity, crumpling phase transitions and nonlinear dynamical structures~\cite{nelson2004statistical}.
While these effects have been initially discussed for  biological membranes, man-made atom-thin membranes from  graphene and other novel two-dimensional crystals have recently  attracted a lot of attention~\cite{Nicholl2013,liu_wu_2016,Khestanova2016,LeDoussal2018}. Understanding  their mechanical behavior is crucial for  charge and thermal~\cite{Balandin2011,Cepelotti2016,Song2018,Kang2018} transport applications. However, despite the ongoing experimental~\cite{Seol2010,Nicholl2013,Ackerman2016} and theoretical~\cite{Gornyi2015,Broido2015,Cepelotti2016} efforts,  the mechanics of the monoatomic membranes remains intriguing. For instance, even the sign of the Poisson ratio for  graphene
is not completely clear~\cite{Burmistrov2018,Kachorovskii2018}. Another family of two-dimensional materials, the transition metal dichalcogenides (TMDCs),  have  strong optical resonances due to the huge exciton binding energy~\cite{Glazov2018} and are, therefore, promising for resonant optomechanics~\cite{Bertolazzi2011,Gomez2013,liu_wu_2016}.

\begin{figure}[b]
\includegraphics[width=0.99\columnwidth]{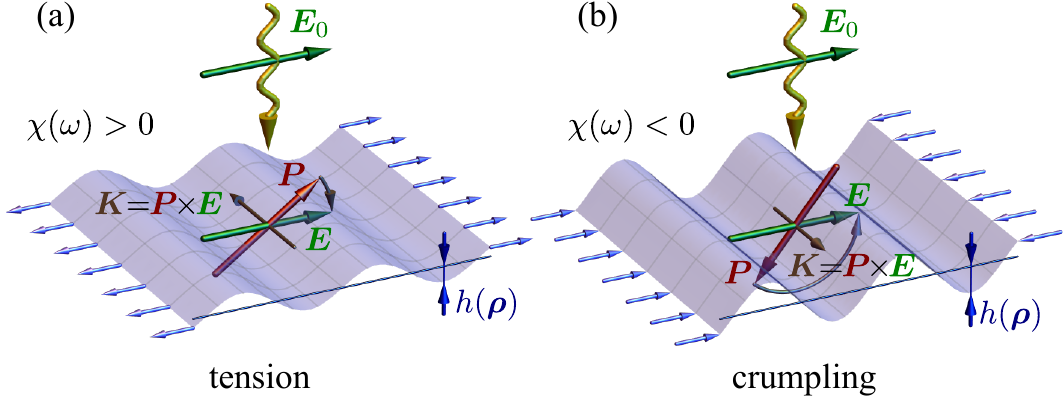}
\caption{ The mechanism of the optomechanical tension. When the deformed membrane is excited by a plane electromagnetic wave, the torque $\bm K = \bm P\times \bm E$  arises and acts upon the membrane. Depending on the sign of the membrane polarizability $\chi(\omega)$, the torque can either (a) flatten or  (b) crumple the membrane. 
}\label{fig:1}
\end{figure}

Here, we predict that mechanical properties of membranes can be strongly affected {by light.}
{We show}
that the optical illumination by a plane wave leads not only to the well-known  pressure of light, but also directly controls the membrane tension.
In a thin membrane,  electric field  induces  polarization 
that is parallel to the membrane surface. When the membrane shape is modulated, the polarization
$\bm P$, induced by the normally incident light, is not aligned with the electric field $\bm E$. This leads to appearance of the torque $\bm K = \bm P \times \bm E$. 
If the membrane polarizability is positive, the torque tends to decrease the amplitude of the modulation, see Fig.~\ref{fig:1}(a). 
This is equivalent to the 
tension that flattens the membrane.
The proposed  optomechanical tension is strongly anisotropic and depends on the light polarization, 
in stark contrast to the well-known radiation pressure and thermally-induced forces~\cite{Akita2017}.
In the opposite case, when the membrane polarizability is negative, the torque tends to increase the amplitude of the {modulation}, see Fig.~\ref{fig:1}(b). 
We demonstrate that such negative optomechanical tension can overcome the intrinsic bending rigidity of the membrane and leads to its transition to the crumpled phase.  
Contrary to the  pearling instability induced by the localized optical tweezers~\cite{BarZiv1994,Barziv1998} and thermal surface structures \cite{Sipe1983,Stratakis2012,Stratakis2014}, the proposed effect takes place under   normal illumination by a homogeneous plane wave and is due to the coherent light-membrane interaction.

\begin{figure}[t]
\includegraphics[width=0.8\columnwidth]{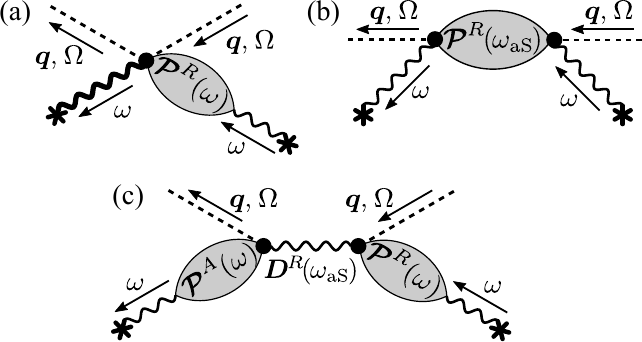}
\caption{The  diagrams describing the self-energy of the flexural vibrations due to (a) second- and (b)--(c) first- order optomechanical interaction with the incident light. Dashed and wavy lines stand for vibrations and light, respectively; bubbles are the polarization operators; dots are the optomechanical interaction vertices $\mathcal{L}_\text{int}^{(1,2)}$. } 
 \label{fig:dia}
\end{figure}

\paragraph{Theory of light-membrane optomechanical interaction.} While the origin of the optomechanical tension is well captured by the na\"\i ve approach illustrated in Fig.~\ref{fig:1}, the rigorous calculation  is highly non-trivial due to  the modification of the electromagnetic field by the membrane itself, i.e., the difference between $\bm E$ and $\bm E_0$ in Fig.~\ref{fig:1}. Such radiative corrections  lead to the  renormalization of the light--membrane interaction, and  become highly important in the most interesting case of strong resonant  polarizability,  realized, e.g., in the TMDC membranes.   
Our aim is to put forward a rigorous theory of the interaction between the electromagnetic field and the flexural vibrations of the optically-resonant membrane and derive the universal expression for  the optomechanical tension via  the light reflection coefficient from the membrane, directly applicable to experiments. 

The straightforward way is to calculate the electromagnetic field by solving the Maxwell equations and accounting for the membrane flection $h$ as a perturbation~\cite{KerkerPRX} and then use the Maxwell stress tensor to determine the linear-in-$h$ force acting on the membrane.  However, such approach turns out to be quite intricate due to the multiple discontinuities and singularities of  the electromagnetic field components at the membrane surface that need to be carefully handled in the stress tensor, see Sec.~S3 of the Supplementary Materials.
Instead, we use the equivalent Lagrangian formalism, that yields the same results in  a significantly more compact fashion and easily handles  all the radiative corrections. To construct the Lagrangian that describes the interaction between the polarization of the deformed membrane and the electromagnetic field, we start from the action $S_\text{int} = \int  \widetilde{\bm{E}} \cdot \widetilde{\bm{P}} \, \rmd\widetilde{\bm r}\, \rmd \widetilde t$, where
$\widetilde{\bm{E}}$, $\widetilde{\bm{P}}$, $\widetilde{\bm r}$ and $\widetilde t$ are the electric field, polarization, coordinate and time, respectively, in the reference frame moving and rotating with the membrane. 
Then, we express the field $\widetilde{\bm{E}}$ through the fields ${\bm{E}}$, $\bm{B}$ in the frame at rest by the subsequent rotation, the spatial shift, and the Lorentz boost
 and obtain
$S_\text{int} = \int \mathcal{L}_\text{int} \rmd t$, where $\rmd t =  (1-\dot{h}^2)^{-1/2} \rmd \widetilde t$, the interaction Lagrangian reads
\begin{align}\label{eq:Lag}
&\mathcal{L}_\text{int} = \int \rmd\bm\rho \, \hat R\widetilde{\bm P}(\bm \rho,t) \cdot \big\{ \bm E_\perp[\bm\rho, h(\bm \rho,t), t] \\ \nonumber
 & + 
(1-\dot{h}^2)^{1/2} \bm e_z E_z[\bm\rho, h(\bm \rho,t), t] + \dot{h}\, \bm e_z \times \bm B[\bm\rho, h(\bm \rho,t), t] \big\} \,,
\end{align}
$\bm e_z$ is a unitary vector along  the undeformed membrane normal $z$, $\bm\rho=(x,y)$ are the in-plane coordinates,  $E_z$ and $\bm E_\perp$ are the electric field components along $z$-axis and perpendicular to it, $\dot{h} = \partial h/c\, \partial t$, and $\hat R$ is the matrix of rotation by the angle $\arctan | \bm\nabla h| $ around the axis $(\partial h/\partial y,-\partial h/\partial x,0)$.
 The Lagrangian Eq.~\eqref{eq:Lag} can be decomposed into the Taylor series over the membrane displacement, $\mathcal{L}_\text{int} = \sum_{m=0}^{\infty} \mathcal{L}_\text{int}^{(m)}$ with $\mathcal{L}_\text{int}^{(m)} \propto h^m$, see the Supplementary Materials for the explicit expressions. The $h$-independent term $\mathcal{L}_\text{int}^{(0)}  = \int d\bm\rho\, \widetilde{ \bm P}(\bm \rho,t) \cdot  \bm E(\bm \rho,t)$ stands for the interaction of light with the undeformed membrane, that we fully take into account. The  remaining terms describe the coupling  between vibrations, polarization and light, and are treated as a perturbation. 
 
\paragraph{Optomechanical  correction to the dispersion of flexural vibrations.}
The optical excitation changes the mechanical properties of the membrane, which is revealed in the modified flexural phonon frequency $\tilde\Omega_{\bm q}$,
\begin{align}\label{eq:disp}
\tilde\Omega_{\bm q}^2 = \Omega_{\bm q}^2 + \Sigma(\bm q,\Omega)/\rho,
\end{align}
where $\Omega_{\bm q} = \sqrt{(\sigma q^2 + \kappa q^4)/\rho}$ is the dispersion of flexural phonons in the absence of optical pump, $\sigma$ is the membrane tension, $\kappa$ is the effective bending rigidity~\cite{Aronovitz1988,Doussal1992,Gornyi2015},  $\rho$ is the membrane mass per unit area, and $\Sigma(\bm q,\Omega)$ is the phonon self-energy correction due to the optomechanical interaction.  To calculate the latter in the non-equillibrium conditions of optical excitation, we exploit the Keldysh Green's function technique~\cite{Ivanov1982,Poshakinskiy2016PRL,Poshakinskiy2017}.
 
In the second order in the optomechanical interaction, the relevant self-energy diagrams are of the three types shown in Fig.~\ref{fig:dia}. 
They include either one vertex corresponding to the second-order interaction $\mathcal{L}_\text{int}^{(2)}$, panel (a), or two vertices  corresponding to the first-order interaction $\mathcal{L}_\text{int}^{(1)}$, panels (b) and (c). The explicit form for $\mathcal{L}_\text{int}^{(1,2)}$ and for the diagrams of  Fig.~\ref{fig:dia} is given in the Supplementary  Sec.~S1. 
Note that we take the light-matter interaction $\mathcal{L}_\text{int}^{(0)}$ in all orders by using the dressed polarization operators (filled bubbles) and the dressed light Green's function (thick wavy line) in contribution (a). In  contributions (b) and (c), the bare light Green's function is used, because the vertex $\mathcal{L}_\text{int}^{(1)}$ is odd under $z$-direction inversion, while the light Green's function is even, and  the dressing vanishes~\cite{KerkerPRX}. 
For the normally incident monochromatic wave with the electric field amplitude $\bm E_0$ and frequency $\omega$, evaluation of the three diagrams in Fig.~\ref{fig:dia} and three more diagrams with the inverted phonon lines yields $\Sigma_\text{om} = \Sigma_\text{aS}(\bm q,\Omega) + \Sigma_\text{S}(\bm q,\Omega)$, where
\begin{align}
&\Sigma_\text{aS}(\bm q,\Omega) = \frac{ | E_{0}|^{2}}{2\pi c}\Bigl\{\omega  \,  {\rm Im\,} r(\omega) \left[ 1 + (cq/\omega)^2 \cos^2\varphi\right]   \nonumber\\
&  
 +\rmi \omega_\text{aS}  \Big[ \frac{r_p(\omega_\text{aS},\theta_\text{aS})}{\cos\theta_\text{aS}}  \cos^2\varphi 
 + r_s(\omega_\text{aS},\theta_\text{aS}) \cos\theta_\text{aS}   \sin^2\varphi \nonumber\\
&\hspace{1.5cm} -|r(\omega)|^2   (1-\sin^2\varphi \sin^2\theta_\text{aS})/\cos\theta_\text{aS}
\Big]\Bigr\} ,
\label{eq:S}
\end{align}
$\Sigma_\text{S}(\bm q,\Omega) = \Sigma_\text{aS}^*(-\bm q,-\Omega)$,  $\omega_\text{aS}=\omega+ \Omega$ is the frequency of the wave anti-Stokes-scattered by the phonon under consideration, $\sin \theta_\text{aS} = q/(\omega + \Omega)$ is the corresponding scattering angle, $r_{s,p}(\omega,\theta) = 2\pi\rmi\omega\chi(\omega)/[c \cos^{\pm1}\theta-2\pi\rmi\omega\chi(\omega)]$
 and $r(\omega) = r_{s,p}(\omega,0)$ are the coefficients of light reflection from the undeformed membrane at the oblique and  normal incidence for two polarizations, 
and $\varphi$ is the angle between $\bm q$ and $\bm E_0$. 

At small $\bm q$ and $\Omega$,  the optomechanical self-energy has the form $\Sigma_\text{om}(\bm q, \Omega) = \sigma_\text{om}(\varphi)\, q^2 -2\rmi \rho \Omega \gamma_\text{om}$. According to Eq.~\eqref{eq:disp}, 
\begin{equation} \label{eq:sigma}
        \sigma_\text{om}(\varphi)=\frac{c|E_0|^2}{\pi\omega} \Im r(\omega) \left[ \cos^2\varphi+\cos 2\varphi \Re r(\omega) \right]
\end{equation}
describes the optomechanical correction to the membrane tension and 
\begin{align}\label{eq:gamma}
\gamma_\text{om} = \frac{|E_0|^{2}}{2\pi\rho c}\left\{ |r(\omega)|^{2} - \Re\frac{\rmd }{\rmd \omega}\left[ \omega r(\omega) \right]\right\}
\end{align}
is the correction to the flexural phonon damping~\footnote{A similar expression for the optomechanical damping rate correction in the system of coupled waveguide and vibrating resonant particle was derived in~\cite{KerkerPRX}}. 

The optomechanical tension, Eq.~\eqref{eq:sigma}, contains two terms of different physical origin{; neither can be derived from the conventional radiation pressure force. }
The first term is due to the torque of the light-induced dipole polarization discussed previously. It is linear in the reflection coefficient; therefore, it dominates when the membrane polarizability is weak and $r\ll 1$.
The second term in Eq.~\eqref{eq:sigma} originates from the Amp\`ere force acting upon the displacement current
in the deformed membrane, see  Sec.~S2 of the Supplementary Materials {for details}. 
Being quadratic in the reflection coefficient, this {term} becomes significant in case of strong reflection, $r \sim 1$, which can  be realized at resonances. 

The obtained optomechanical tension Eq.~\eqref{eq:sigma}, induced by the linearly polarized light, is strongly anisotropic.  
The tension {along}
($\varphi = 0$) and perpendicular ($\varphi = \pi/2$) to the electric field of the incident light is given by
 \begin{align}
 \sigma_{\text{om}}^\parallel = \frac{c|E_0|^2}{\pi\omega} \Im r \Re t , \quad \sigma_{\text{om}}^\perp = - \frac{c|E_0|^2}{\pi\omega}  \Im r \Re r \,,
 \end{align}
where $t  = 1+r$ is the transmission coefficient. We note that $ 0 \leq \text{Re\,}t \leq 1$ and $ -1 \leq \text{Re\,}r \leq 0$  in the absence of gain. 
Therefore, the sign of both tension components is determined by the sign of $\Im r$. In case of unpolarized or circularly polarized excitation, the optomechanical tension is isotropic and given by $\langle \sigma_{\text{om}}\rangle = (\sigma_{\text{om}}^\parallel+\sigma_{\text{om}}^\perp)/2$.

\begin{figure}[t]
\includegraphics[width=0.99\columnwidth]{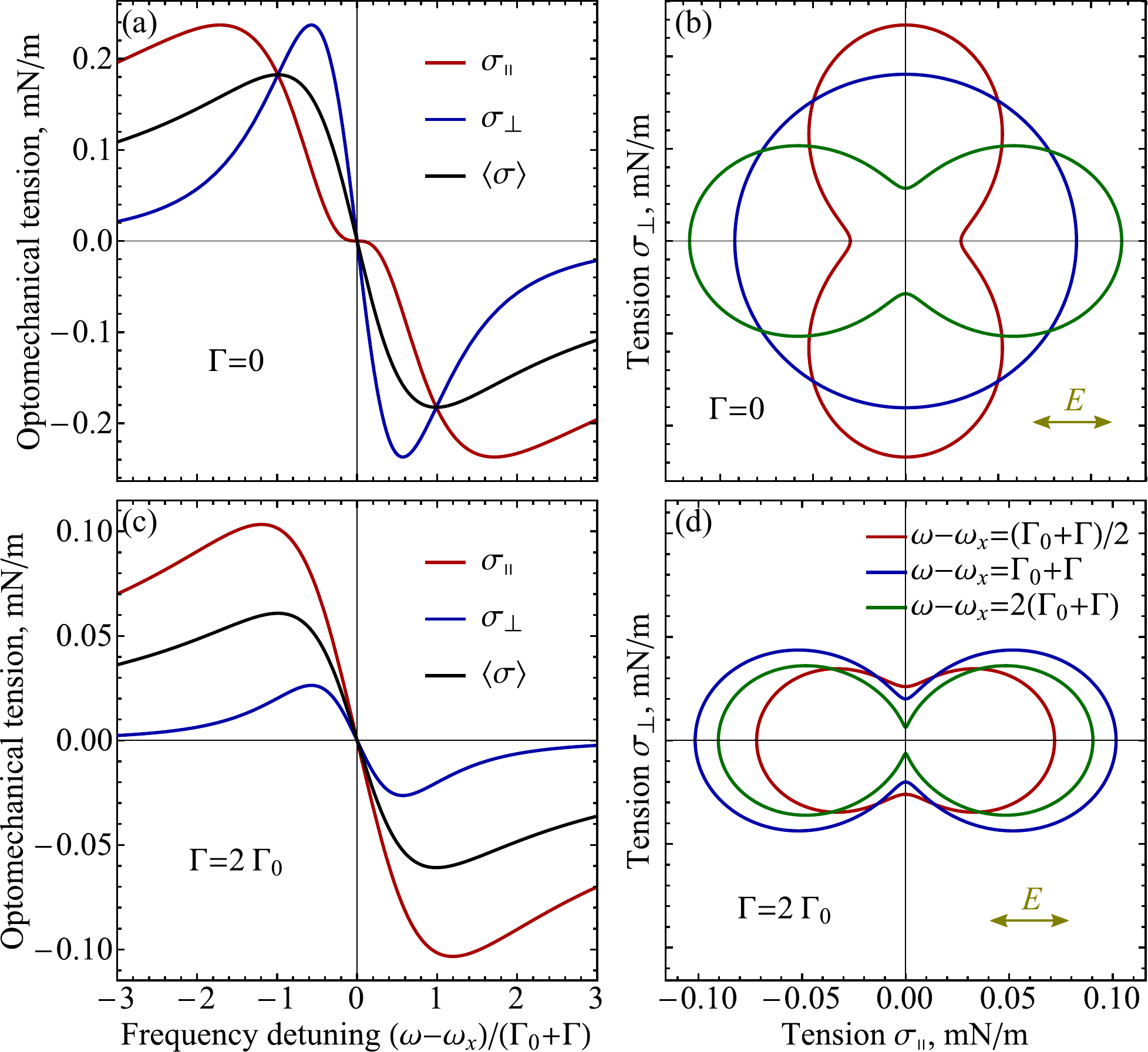}
\caption{  (a),(c) Optomechanical tension along ($\sigma_{\text{om}}^\parallel $) and perpendicular ($\sigma_{\text{om}}^\perp$) to the electric field of the incident linearly polarized light and the tension $\langle \sigma_{\text{om}}\rangle$ induced by the unpolarized light depending on the detuning of the light frequency  from the resonance. (b),(d) Dependence of the optomechanical tension on the in-plane direction for linearly polarized excitation and three different values of the frequency detuning. Calculation is made for the 1\,W/$\mu$m$^2$ light intensity; nonradiative broadening $\Gamma=0$ for panels (a),(b) and  $\Gamma=2\Gamma_0$ for panels (c),(d). 
}\label{fig:3}
\end{figure}


   \paragraph{Frequency and polarization dependence.} 
To be specific, we consider the membrane with the reflection coefficient of the form 
    \begin{equation}
   r(\omega)=\frac{\rmi \Gamma_{0}}{\omega_x-\omega-\rmi (\Gamma_{0}+\Gamma)}\:,\label{eq:r1}
   \end{equation}
 which is realized, e.g., for the TDMC membranes in the vicinity of the exciton resonance frequency $\omega_{x}$. Here, $\Gamma_{0}$ and $\Gamma$ are the  radiative and nonradiative broadening of the resonance, respectively.  The latter is governed by the quality of the structure~\cite{Robert2016}.  Figure~\ref{fig:3}(a) shows the frequency dependence of the optomechanical tension components for the resonant membrane with small losses, $\Gamma< \Gamma_0$. 
The tension is positive below the resonance and negative above it. 
Interestingly, in the vicinity of the resonance, $|\omega-\omega_x|<\Delta^* =\sqrt{\Gamma_0^2-\Gamma^2}$, the perpendicular tension 
is larger than the parallel one, for $|\omega-\omega_x|=\Delta^*$ the tension is isotropic, while for  $|\omega-\omega_x|>\Delta^*$ the parallel tension dominates.
{The angular dependence of the tension for these three cases is shown in Fig.~\ref{fig:3}(b)}.
The case of resonance with significant losses, $\Gamma > \Gamma_0$ is illustrated in Figs.~\ref{fig:3}(c),(d). For all detunings, the optomechanical tension is predominantly along the electric field. Indeed, in such case the reflection coefficient is small and the optomechanical tension is determined by the first term of Eq.~\eqref{eq:sigma}. 

For the light with intensity of 1\,W/$\mu$m$^2$, we estimate that the optomechanical tension can reach $0.2$\,mN/m. Such a correction to the membrane tension shall lead to a measurable change of the vibration eigenfrequencies for a typical suspended membrane {with mechanical quality factor $Q \gtrsim 10^3$ pretensioned to}  $\sigma \sim 10\,\rm mN/m$~\cite{Bertolazzi2011,liu_wu_2016,Gomez2013}.
 The characteristic polarization and frequency dependence of the optomechanical tension can be used to distinguish it from the radiation pressure and heating due to the light absorption. 
  
\begin{figure}[t]
\includegraphics[width=0.99\columnwidth]{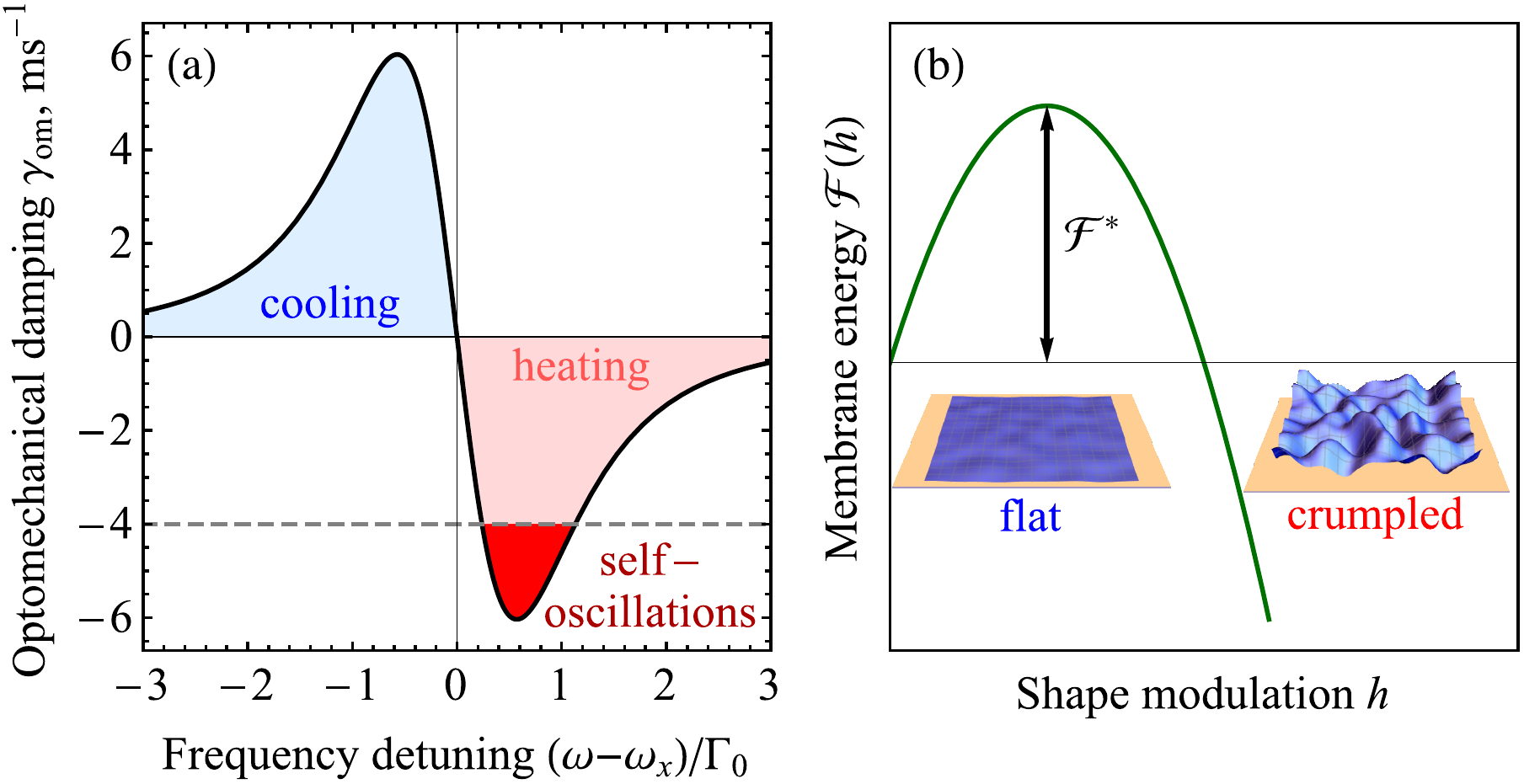}
\caption{Optomechanical instabilities. (a) Optomechanical correction to the damping rate of flexural vibrations as a function of the light frequency detuning from the resonance.  Calculation is performed for the light intensity 1\,W/$\mu$m$^2$, $\rho=6.4\times 10^{-7}\,\rm g/cm^{2}$, and exciton resonance parameters, $\hbar\omega_x=1.6$\,eV, $\hbar\Gamma_{0}=0.3$\,meV~\cite{Robert2016}, $\Gamma=0$. When the negative optomechanical correction to the damping overcomes the intrinsic phonon decay rate, indicated by the dashed curve for a lateral membrane size $L=20\,\mu$m, the conditions for  self-oscillations are realized. 
%
(b) Dependence of the membrane energy on the amplitude of the membrane shape modulation featuring the barrier that separates the flat and the crumpled states. At temperatures higher than $\mathcal{F}^*$, transition to the crumpled phase occurs. 
}\label{fig:4}
\end{figure}

Finally, we analyze the optomechanical correction to the decay rate of flexural vibrations in resonant membranes. In the vicinity of the resonance, the reflection coefficient $r$ changes rapidly and Eq.~\eqref{eq:gamma} simplifies to   
  $\gamma_\text{om} = - \omega |E_0|^2 \Re(  \rmd r / \rmd\omega) /(2\pi\rho c) $.
For the  resonant reflection coefficient {of the form of} Eq.~\eqref{eq:r1}, 
the correction to the  decay rate {is} shown in Fig.~\ref{fig:4}(a).
Under the excitation with frequency below the resonance, the damping of the membrane flexural vibrations is increased,  while it is suppressed when the frequency is above the resonance. This agrees  with the concept of optomechanical  cooling and heating effects~\cite{Kippenberg2014,KerkerPRX}. 

\paragraph{Optomechanical instabilities.} When the optomechanical tension or the optomechanical damping are negative, the flat membrane  can become unstable. First, we discuss the instability caused by the optomechanical heating, $\gamma_\text{om}<0$. 
 The intrinsic phonon decay rate decreases for small  phonon frequencies $\Omega$, while $\gamma_\text{om}$ remains constant. Hence, the total  damping of low-frequency vibrations is negative. The  fluctuations of the membrane shape with the corresponding wave vectors are amplified, promoting  a self-oscillatory or chaotic behavior~\cite{Kippenberg2014}. 
Assuming that the low-frequency vibrations in the free membrane have the intrinsic quality factor of the order of unity due to the strong anharmonicity~\cite{Broido2015,BurKach}, we estimate that, under 1\,W/$\mu$m$^2$ excitation, such an instability is realized for the phonons with wave vectors $q \lesssim 0.2\,\mu$m$^{-1}$, which requires the membrane with the lateral dimensions $L \gtrsim 20\,\mu$m.
    
Now we discuss the possibility of the {\it crumpling instability} caused by $\sigma_\text{om}<0$.
To crumple the membrane in a typical suspended mechanical resonator, one should overcome the pretension  $\sigma \sim 10~\rm mN/m$ by the negative optomechanical tension. This requires the light intensity $\sim 30\,\rm W/\mu m^{2}$  that seems to be reasonably below the optical damage threshold in the regime of pulsed excitation~\cite{Roberts2011,Stratakis2014}.
 However, this consideration ignores the effect of the pressure of light. In the limit of small deformations, the deflection induced by the pressure of light can be estimated as   $h_0\sim |E|^2 L^{2}/\sigma$. For the \mbox{30-$\rm W/\mu m^{2}$} excitation and the lateral size $L\sim 1~\rm \mu m$, we find  $h_0\sim 10~\mu {\rm m}$, which is clearly beyond the small flexure limit.  This means that the membrane shape is controlled by the pressure of light directly, rather than by the proposed optomechanical tension. The undesired light pressure  can be suppressed by placing the membrane in an antinode of a standing electromagnetic wave in a cavity or just by considering smaller  flakes.

 An  alternative to the suspended pretensioned membranes is provided by the membranes  freely lying on a substrate. 
 In this case the membrane shape is determined by a competition of the optomechanical crumpling, the radiation pressure, and the membrane rigidity. The potential energy change due to the modulation of the membrane shape with the amplitude $h$ and the wave vector $q$ can be estimated as
 \begin{align}
 \mathcal{F}= (\sigma_\text{om} q^2 h^2 + \kappa q^4 h^2 + 2 p |h|) L^2 \,
 \end{align}
 where $ p = -\text{Re\,}r\, |E_0|^2/\pi $ is the radiation pressure,  $\kappa \sim 1\,$eV is the membrane bending rigidity, and  we assume $\sigma_\text{om}<0 $. For the wave vectors $q <\sqrt{-\sigma_\text{om}/\kappa}$ the dependence $ \mathcal{F}(h)$ is sketched in Fig.~\ref{fig:4}(b). While at large $h$ the optomechanical crumpling dominates, there is a light-pressure barrier of the height $ \mathcal{F}^* = -L^2p^2/(\sigma_\text{om} q^2  + \kappa q^4)$ that separates the flat  ($h=0$) and crumpled ($h \to \infty$) phases. The minimal barrier  $ 4Sp^2 \kappa/\sigma_\text{om}^2$  is realized for the wave vector $q^* = \sqrt{-\sigma_\text{om}/2\kappa}$. Considering the smallest membrane that supports such a modulation, $L = \pi/q^*$, we finally obtain
 \begin{align}
 \mathcal{F}^*_\text{min} = -\frac{8\pi^2 p^2 \kappa^2}{\sigma_\text{om}^3}  =  -\frac{64\pi^3 \omega^3 \kappa^2}{c^3 |E_0|^2} \, \frac{(\text{Re}\,r)^2}{(\text{Im}\,r)^3} \,
 \end{align}
 for unpolarized light. 
The barrier height is inversely proportional to the light intensity.
When $ \mathcal{F}^*_\text{min} $ becomes comparable with the temperature, the thermal fluctuations push the membrane over the barrier and the  transition to the crumpled phase occurs. 
In the resonant conditions, we estimate the intensity required for the crumpling at room temperature as 0.5\,W/$\mu$m$^2$ for $\hbar\omega = 0.1\,$eV.  The moderate detuning from the resonance, which leads to a decrease of $(\text{Re}\,r)^2/(\text{Im}\,r)^3 \sim 1/(\omega-\omega_x)$, can be used to further facilitate the crumpling. For the polarized excitation, the barrier height depends on the wave vector direction. In the vicinity of the resonance (away from the resonance), the crumpling starts in the direction perpendicular (parallel) to the electric field, see Fig.~\ref{fig:3}.

   Our theoretical findings indicate the huge potential of membranes for the fusion of nonlinear mechanics with the resonant nano-optics.  While the optomechanics of non-resonant membranes is already a mature field with recent  successful demonstration of the spontaneous symmetry-breaking buckling of a membrane as a whole~\cite{Buchmann2012,Xu2017}, proposed optomechanical crumpling harnesses the non-homogeneous membrane deformations. Even though our consideration is oversimplified by ignoring the strongly non-Hookean elasticity of the membranes~\cite{Gornyi2015} as well as the effect of adhesion~\cite{Khestanova2016}, in our opinion it clearly indicates that the optomechanical tension and crumpling is potentially important for  realistic membranes. 
   The effect could be also enhanced when the wave vector of the incident or scattered light is in  resonance with the surface waves, supported by the membrane.  
   
   Moreover, our results are not limited to graphene and TMDC membranes in the optical frequency range.  Since the optomechanical tension Eq.~\eqref{eq:sigma} is inversely proportional to the light wavelength, one could expect interesting physics at the lower frequencies, in the THz or even radio frequency spectral range.  Optomechanical tension can be used  for visualization of fundamental physics of phase transitions and effective mechanical field theories, tailoring the heat and charge transport as well as  modification of the light-matter coupling at the nanoscale~\cite{Camerer2011,Vochezer2018}. Other potential  platforms for  the proposed effect range from the    biological membranes, where the tensions of 
 $10^{-5}\rm N/m$ are routinely studied by optical tweezers ~\cite{HOCHMUTH1996,Pontes2017} to the   structured solar sails, now attracting a lot of attention~\cite{Atwater2018,Ilic2019}. 
   
  \tocless\acknowledgements
  We acknowledge useful discussions with  I.S. Burmistrov,  V.Yu. Kachorovskii, V. Menon and T. Smolenski.
  This work has been supported by the Russian Science Foundation Grant No. 19-12-00051. The authors acknowledge partial support from the Russian President Grant No. MD-5791.2018.2 and the Foundation for the Advancement of Theoretical Physics and Mathematics ``BASIS''. A.V.P. also acknowledges partial support from the Russian President Grant No. MK-599.2019.2.


%


\onecolumngrid

\setcounter{equation}{0}
\newcounter{sfigure}
\setcounter{sfigure}{1}
\setcounter{table}{0}
\makeatletter
\renewcommand{\theequation}{S\arabic{equation}}

 \renewcommand\thefigure{S{\arabic{sfigure}}}
\renewcommand{\thesection}{S\arabic{section}}
\renewcommand{\thesubsection}{S\arabic{subsection}}

\vspace{.02\paperheight}
\newpage

\begin{center}
\textbf{\large Supplemental Material\\ for ``Optomechanical tension and crumpling of resonant membranes''}
\end{center}
\tableofcontents
\section{S1. Optomechanical self-energy for flexural vibrations}

\subsection{First- and second-order optomechanical interaction vertices}

Decomposing the Lagrangian~\eqref{eq:Lag} into the Taylor series over the membrane displacement $h(\bm\rho)$, we obtain
the linear-in-$h$ optomechanical interaction
\begin{align}
&\mathcal{L}_\text{int}^{(1)} =  \int 
\bm P \cdot \left[h \frac{\partial \bm  E}{\partial z} +  \dot{h} (\bm e_z \times \bm B) -\bm s \times \bm B 
\right]   d\bm\rho \,,
\end{align}
and the $h$-quadratic interaction
\begin{align}
&\mathcal{L}_\text{int}^{(2)} =  \int 
\bm P \cdot \left[\frac{h^2}{2} \frac{\partial^2 \bm  E}{\partial z^2} + h\dot{h} \, \bm e_z \times  \frac{\partial \bm B}{\partial z} 
-h \bm s \times \frac{\partial \bm E}{\partial z} - \dot{h} (\bm s \cdot \bm B) \bm e_z + \bm s \times (\bm s \times \bm E) - \frac{\dot{h}^2}2 E_z \bm e_z
\right]   d\bm\rho \,,
\end{align}
where $\bm s = (\partial h/\partial y, -\partial h/\partial x, 0)$. 
Next, we express the optomechanical action $S_\text{om}^{(1,2)} = \int \mathcal{L}_\text{int}^{(1,2)}$ in terms of  the vector potential and the current determined by $\bm E = -(\partial \bm A/ \partial \bm t)$ and $\bm j = \partial \bm P/ \partial \bm t$, and switch to the Fourier space. We obtain then $S_\text{om}^{(1,2)} = -\bm P(\omega') \cdot \bm \Lambda^{(1,2)}(\omega',\omega) \bm A(\omega)$, 
where the first-order~\cite{KerkerPRX}  and the second-order optomechanical vertices read
\begin{align} \label{eq:s:L1}
&\bm \Lambda^{(1)} (\omega',\Omega_1,\omega)= \rmi h_1 \left[ k_z -\frac{\Omega_1}{\omega'} \bm k \otimes \bm e_z - \frac{\omega}{\omega'}(\bm e_z \otimes \bm q -\bm q \otimes \bm e_z)\right]\:, \\ 
&\bm \Lambda^{(2)} (\omega',\Omega_1,\Omega_2,\omega)
=  h_1 h_2 \left( -k_z^2   + \frac{\Omega_1+\Omega_2}{\omega'} k_z  \bm k \otimes \bm e_z + \frac{\omega\Omega_1\Omega_2}{\omega'}  \bm e_z \otimes \bm e_z \right) 
  + \frac{\omega}{\omega'} \left( \frac{\bm s_1 \otimes \bm s_2 + \bm s_2 \otimes \bm s_1}2-\bm s_1 \cdot \bm s_2 \right)\:,
\nonumber\\ \label{eq:s:L2}
& -\frac{\rmi h_1 \bm s_2}{\omega'} \times \left[ (\omega+\Omega_1) k_z - \Omega_1 \bm k \otimes \bm e_z\right]
 -\frac{\rmi h_2 \bm s_1}{\omega'} \times \left[ (\omega+\Omega_2) k_z - \Omega_2 \bm k \otimes \bm e_z\right] \:,
\end{align}
where $h_{1,2}$, $\bm q_{1,2}$, and $\Omega_{1,2}$ are the amplitudes, wave vectors, and frequencies of the involved vibrations, $\bm s_{1,2} = \rmi h_{1,2 }\bm q_{1,2} \times \bm e_z$,  $\bm k$ is the wave vector corresponding to $\bm A$, and $\omega' = \omega+\Omega_1 (+\Omega_2)$ is assumed for the first(second)-order vertex.

\subsection{Evaluation of the optomechanical self-energy correction}

We consider here the general case when the pump wave at frequency $\omega$ is incident at the angle $\theta$ with respect to the membrane normal. For clarity, we set $\hbar = c = 1$. 

The self-energy correction to the flexural phonon dispersion is given by the three diagrams depicted in Fig.~\ref{fig:dia},
\begin{align}
&\Sigma_\text{aS,out} = -[\bm\Lambda^{(1)}(\omega_{\rm aS},\Omega,\omega)  \bm A]^* \cdot \bm{\mathcal{P}}(\omega_{\rm aS}) \bm\Lambda^{(1)}(\omega_{\rm aS},\Omega,\omega) \bm A ,\\
&\Sigma_\text{aS,in} =-   [{\bm\Lambda^{(1)}}^T(-\omega,\Omega,-\omega_{\rm aS}) \bm{\mathcal{P}}(\omega)\bm A]^* \cdot \bm{D}(\omega_{\rm aS}) {\bm\Lambda^{(1)}}^T(-\omega,\Omega,-\omega_{\rm aS}) \bm{\mathcal{P}}(\omega)\bm A ,\\
&\Sigma_\text{aS,2} =- [\bm{\mathcal{P}}(\omega)\bm A]^* \cdot  \bm\Lambda^{(2)} (\omega,-\Omega,\Omega,\omega) \bm{\mathcal{A}},
\end{align}
and three more diagrams with the inverted direction of pump laser lines, which correspond to Stokes scattering and yield
\begin{align}
& \Sigma_\text{S,out} = -[\bm\Lambda^{(1)}(\omega_{\rm S},-\Omega,\omega)  \bm A]^* \cdot \bm{\mathcal{P}}^*(\omega_{\rm S}) \bm\Lambda^{(1)}(\omega_{\rm S},-\Omega,\omega)  \bm A ,\\
&\Sigma_\text{S,in} =-[{\bm\Lambda^{(1)}}^T(-\omega,-\Omega,-\omega_{\rm S}) \bm{\mathcal{P}}(\omega)\bm A]^* \cdot \bm{D}^*(\omega_{\rm S}) {\bm\Lambda^{(1)}}^T(-\omega,-\Omega,-\omega_{\rm S}) \bm{\mathcal{P}}(\omega)\bm A  ,\\
&\Sigma_\text{S,2} =- \bm{\mathcal{A}}^* \cdot  {\bm\Lambda^{(2)}}^T (\omega,-\Omega,\Omega,\omega) \bm{\mathcal{P}}(\omega)\bm A  ,
\end{align}
where $\bm A = A_s\bm e_s + A_p \bm e_p$ is the vector potential of the incident light, $\bm e_s$ and $\bm e_p$ are the unitary vectors representing $s$- and $p$-polarization, $\bm{\mathcal{A}}= (1+\bm{D\mathcal P}) \bm A$ is the vector potential dressed by the interaction with layer polarization, $\bm{\mathcal{P}}$ is the dressed polarization operator of the layer, 
\begin{align}\label{eq:s:P}
\bm{\mathcal{P}}_{\bm k}(\omega) = \mathcal P_t \left( 1 - \frac{\bm k_{\perp} \otimes \bm k_{\perp}}{k_{\perp}^2}\right)  
+ \mathcal P_l\,  \frac{\bm k_{\perp} \otimes \bm k_{\perp}}{k_{\perp}^2}  \, \,, \quad
\mathcal P_l = \frac{\omega}{2\pi\rmi} \,\frac{r_p(\omega,\theta)}{\cos\theta} \,,\quad
\mathcal P_t = \frac{\omega}{2\pi\rmi} \,r_s(\omega,\theta) {\cos\theta}
\end{align}
with $\bm k_{\perp} = (k_x, k_y)$ being the in-plane wave vector and $\cos \theta = \sqrt{\omega^2-k^2_{\perp}}/ k_{\perp}$,
$\bm D$ is retarded Green's function of the bare photon
\begin{align}\label{eq:s:G}
\bm D_{\bm k}(\omega) = -\frac{4\pi}{(\omega+ \rmi 0)^2- \bm k^2} \, \left( 1 - \frac{\bm k \otimes \bm k}{\omega^2} \right) \,,
\end{align}
$\omega_{\rm aS(S)} = \omega \pm \Omega$ is the frequency of anti-Stokes(Stokes)-scattered wave. 

\subsubsection{Calculation of $\Sigma_\mathrm{aS,out}$, diagram Fig.~\ref{fig:dia}(b)}

\begin{figure}[t]
\includegraphics[width=0.25\textwidth]{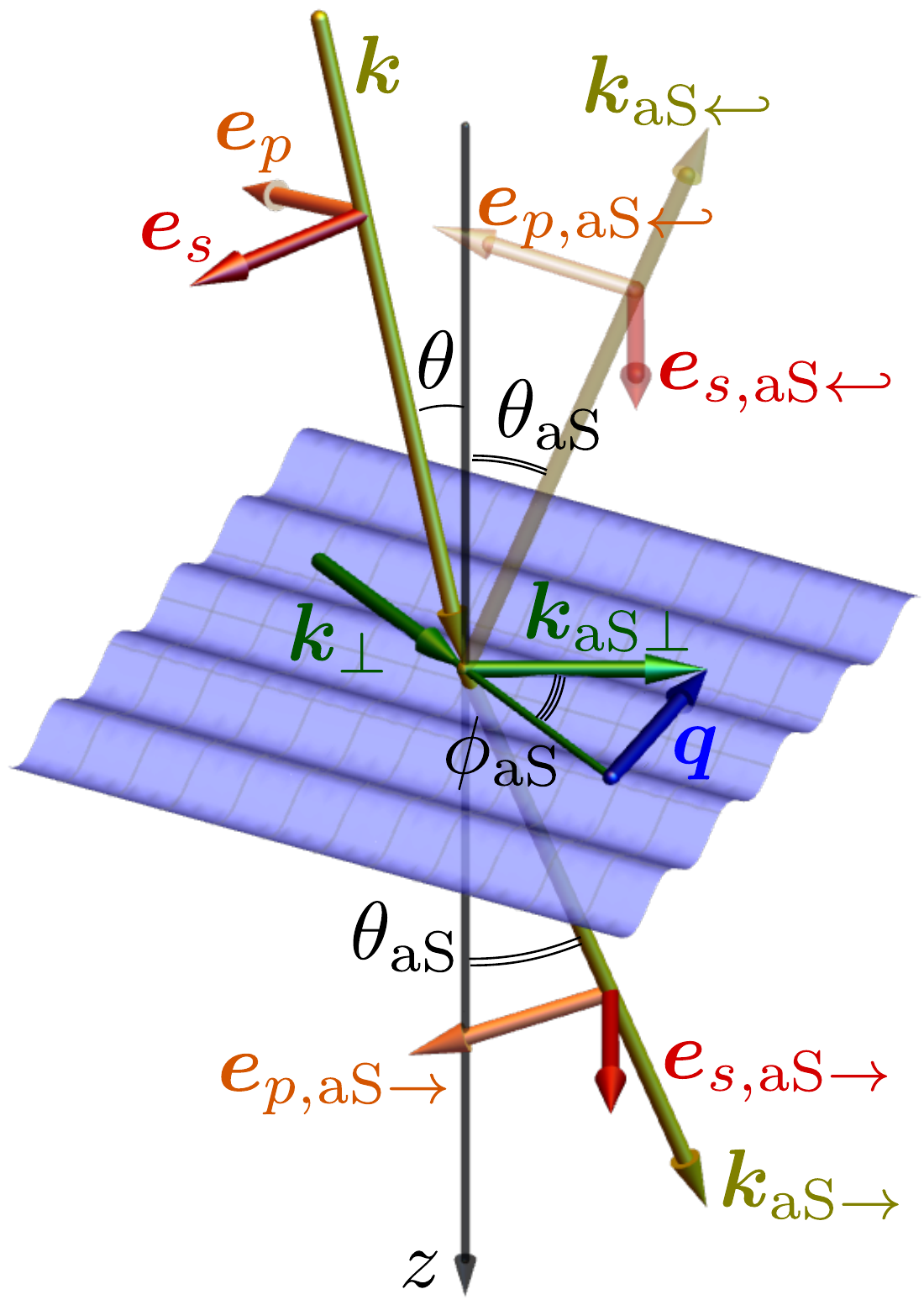}
\caption{A sketch of the anti-Stokes forward and backward light scattering on the flexural vibration of the membrane, that introduces the notations for scattering angles and polarization vectors. }\label{fig:S:coors}
\stepcounter{sfigure}
\end{figure}

First, we calculate the action of the first-order optomechanical interaction operator on the incident field, 
\begin{align}
\bm\Lambda^{(1)} \bm A = \rmi \left[ k_z \bm A +\left(\frac{\omega}{\omega'} \bm k_{\text{aS}\perp}- \bm k_{\perp}\right) A_z \right]  + (\ldots)\, \bm e_z
\end{align}
where $\bm k_{\text{aS}\perp} = {\bm k}_{\perp} + \bm q$ and  the ellipsis stands for the out-of-plane contribution that does not enter the subsequent calculations. Then, we decompose $\bm\Lambda^{(1)} \bm A$ into the components parallel and perpendicular to $\bm k_{\text{aS} \perp} $,  $\bm\Lambda^{(1)} \bm A = (\bm\Lambda^{(1)} \bm A)_{l'} \bm e_{l'}  + (\bm\Lambda^{(1)} \bm A)_{t'} \bm e_{t'}$. The components read
\begin{align}
(\bm\Lambda^{(1)} \bm A)_{l'} &= \rmi \omega  \left[ -A_s \cos \theta \sin\phi_\text{aS}  - A_p (\cos\phi_\text{aS}-\sin\theta\sin\theta_\text{aS}) \right] ,\nonumber\\
(\bm\Lambda^{(1)} \bm A)_{t'} &= \rmi \omega  \left[ A_s \cos \theta \cos\phi_\text{aS}  - A_p \sin\phi_\text{aS} \right] ,
\end{align}
where $\theta_\text{aS} = \arcsin k_{\text{aS} \perp}/\omega_{\text{aS}}$ and $\phi_\text{aS}$ is the angle between $\bm k_\perp$ and $\bm k_{\text{aS} \perp}$, see Fig.~\ref{fig:S:coors}, and the basis of $\bm e_{l'}$ and $\bm e_{t'}$ is determined in such way that $\bm e_t = \bm e_{s,\text{aS}}$ and $(\bm e_l \times \bm e_t)\cdot \bm e_z >0$.  
Then, we apply the $\bm{\mathcal{P}}_{\bm k_\text{aS}}(\omega_\text{aS})$ polarization operator, Eq.~\eqref{eq:s:P},
and find
\begin{align}\label{eq:s:Sout}
\Sigma_\text{aS,out} =  \frac{\rmi \omega^2\omega_{\rm aS} }{2\pi}  
\Big[ r_s(\omega_{\rm aS},\theta_{\rm aS}) \cos\theta_{\rm aS} \Big| A_s \cos \theta \cos\phi_{\rm aS} - A_p \sin\phi_{\rm aS} \Big|^2 \nonumber \\
+ \frac{r_p(\omega_{\rm aS},\theta_{\rm aS})}{\cos\theta_{\rm aS}}\Big| A_s \cos \theta \sin\phi_{\rm aS}  + A_p (\cos\phi_{\rm aS}-\sin\theta\sin\theta_{\rm aS}) \Big|^2 \Big] \,.
\end{align}

\subsubsection{Calculation of $\Sigma_\mathrm{aS,in}$, diagram Fig.~\ref{fig:dia}(c)}

First, we calculate the polarization current induced by the incident wave,
\begin{align}
&\bm j = \bm{\mathcal{P}}\bm A = \frac{\omega}{2\pi\rmi} (r_s \cos\theta A_s \bm e_t - r_p A_p \bm e_l)\:.
\end{align}
Application of the first-order optomechanical vertex yields
\begin{align}
{\bm\Lambda^{(1)}}^T \bm j = k_z' \bm j  - \left[ \left( \bm k' - \frac{\omega_\text{aS}}{\omega} \bm k \right) \cdot \bm j \right] \bm e_z\:.
\end{align}
Then, using the Green function $\bm D_{\bm k'}(\omega_\text{aS})$, Eq.~\eqref{eq:s:G}, we obtain
\begin{align}
& (\bm\Lambda^T \bm j)^* \cdot \bm D^R (\bm\Lambda^T \bm j) =-\frac{4\pi}{(\omega+ \rmi 0)^2- \bm k^2} \left\{\left| \left( \bm k' - \frac{\omega_\text{aS}}{\omega} \bm k \right) \cdot \bm j \right|^2
+ k_z'^2 \left( |\bm j|^2 - \frac{|\bm k \cdot \bm j|^2}{\omega^2}\right) \right\} \,,
\end{align}
which has to be summed over $k_z'$ yielding
\begin{align}
 2K \left( |\bm j|^2 - \frac{|\bm k \cdot \bm j|^2}{\omega^2}\right) + \frac{2\pi\rmi}{k_z'} \left[ k_z'^2\left( |\bm j|^2 - \frac{|\bm k \cdot \bm j|^2}{\omega^2} \right) + \left| \left( \bm k' - \frac{\omega_\text{aS}}{\omega} \bm k \right) \cdot \bm j \right|^2 \right] \,,
\end{align}
where now $k_z' = \sqrt{\omega_\text{aS}^2-\bm k_{\text{aS}\perp}^2}$ and $K = \int dk_z $ is a diverging constant.  
Finally, we get the self energy correction
\begin{align}
&\Sigma_\text{aS,in} = -2K \left( \frac{\omega}{2\pi}\right)^2 \cos^2\theta \left( |r_s (\omega,\theta)A_s|^2 +|r_p(\omega,\theta) A_p|^2 \right) 
- \rmi \frac{\omega^2 \omega_{\rm aS}}{2\pi} \cos\theta_{\rm aS} \\
&\times \left[ \cos^2\theta \left( |r_s(\omega,\theta) A_s|^2 +|r_p(\omega,\theta) A_p|^2 
\right)+ \frac{\left| \cos\theta\sin\theta_{\rm aS}\sin\phi_{\rm aS} r_s(\omega,\theta) A_s - (\sin\theta-\sin\theta_{\rm aS}\cos\phi_{\rm aS})r_p(\omega,\theta) A_p\right|^2}{\cos^2\theta_{\rm aS}} \right] \,,
\nonumber
\end{align}
which can be alternatively rewritten as
\begin{align}\label{eq:s:Sin}
&\Sigma_\text{aS,in} = -2K  \left( \frac{\omega}{2\pi}\right)^2 \cos^2\theta \left(|r_s(\omega,\theta) A_s|^2 +|r_p (\omega,\theta)A_p|^2 \right) 
- \rmi  \frac{\omega^2 \omega_{\rm aS}}{2\pi} \cos\theta_{\rm aS} \\
&\times \left[ 
\Big|
\cos\theta\cos\phi_{\rm aS} r_s(\omega,\theta) A_s + \sin\phi_{\rm aS} r_p(\omega,\theta) A_p
\Big|^2+
\frac{\Big|\cos\theta\sin\phi_{\rm aS} r_s(\omega,\theta) A_s + (\cos\phi_{\rm aS} - \sin\theta\sin\theta_{\rm aS}) r_p(\omega,\theta) A_p
\Big|^2}{\cos^2\theta_{\rm aS}}
\right] \,.
\nonumber
\end{align}

\subsubsection{Calculation of $\Sigma_\mathrm{aS,2}$, diagram Fig.~\ref{fig:dia}(a)}

For $\bm h_1 = \bm h_2^*$ and $\Omega_1=-\Omega_2 = \Omega$, we get $\bm s_1 = \bm s_2^* = \rmi h_1 [\bm q \times \bm e_z]$ and the second-order optomechanical interaction vertex~Eq.~\eqref{eq:s:L2} reduces to
\begin{align}
\Lambda^{(2)} = \left( \frac{\partial^2}{\partial z^2}  - \bm q \otimes \bm q \right) + \ldots
\end{align}
where the ellipsis replaces the out-of plane components that do not enter the following calculations. Then, we obtain
\begin{align}\label{eq:s:S2}
\text{Re\,}\Sigma_\text{aS,2} = \frac{2\omega^2\cos^2\theta\, K}{(2\pi)^2} &\left( |r_s(\omega,\theta) A_s|^2 + |r_p(\omega,\theta)A_p|^2\right) + \frac{\omega^3 \cos^3\theta}{2\pi}\text{Im\,} \left( r_s(\omega,\theta) |A_s|^2 + r_p(\omega,\theta) |A_p|^2 \right) \nonumber \\
+ \frac{\omega q^2}{2\pi} \text{Im} &\left[\cos\theta \left( r_s(\omega,\theta) |A_s|^2 \sin^2\alpha +  r_p(\omega,\theta) |A_p|^2 \cos^2\alpha\right) \right. \nonumber \\
&\left.+A_s^* A_p  \left( r_s^*(\omega,\theta) \cos\theta^2  - r_p(\omega,\theta) - r_s^*(\omega,\theta) r_p(\omega,\theta) \sin^2 \theta \right) \sin\alpha\cos\alpha \right]
\end{align}
where we introduced   $\bm q \cdot \bm e_l = q \cos\alpha$ and $\bm q \cdot \bm e_t = q \sin\alpha$. The imaginary part of $\Sigma_\text{aS,2}$ is irrelevant, since it is cancelled by $\text{Im\,}\Sigma_\text{S,2} = -\text{Im\,}\Sigma_\text{aS,2}$.  Note that the singular term $\sim K$ cancels out when the contributions $\Sigma_\text{in,aS}$ and $\Sigma_\text{aS,2}$ are summed. 

When the normal incidence is considered, $\theta = 0$, the summation of the three contributions, $\Sigma_\text{aS,out}$, $\Sigma_\text{aS,2}$ and $\Sigma_\text{aS,2}$, Eqs.~\eqref{eq:s:Sout}, \eqref{eq:s:Sin}, and~\eqref{eq:s:S2}, yields Eq.~\eqref{eq:S}
 of the main text.

\section{S2. The physical origins of the optomechanical tension} 
 In this section we present an alternative  derivation of the optomechanical tension, Eq.~\eqref{eq:sigma}.
 The goal is to clarify  the microscopic origin of the two contributions. To this end, we consider a static modulation of the membrane shape, $\Omega=0$, and calculate the optical forces that appear. 
  We stress that the optomechanical tension force cannot be derived from the conventional radiation pressure force. Indeed, the latter is a function of the local membrane tilt angle $ \bm\nabla h$, while the former should be proportional to $\nabla^2 h$.

The first term in Eq.~\eqref{eq:sigma} is due to the torque acting upon the membrane polarization, as illustrated in Fig.~\ref{fig:1}. The dipole polarization  induced by normally incident light in the flat membrane reads $\bm P = \chi \bm E$, where $\bm E$ is electric field in the membrane. Membrane flexion leads  to an appearance of the $z$ component of the polarization, proportional to the tilt,
\begin{align}
P_z = \frac{\partial h}{\partial \bm \rho} \cdot \bm P \,.
\end{align}
The torque acting on the polarization, $\bm K  = 2\,\text{Re}( \bm P \times \bm E^* )$, is spatially inhomogeneous, and leads to the force
\begin{align}
f_z = \left(\frac{\partial}{\partial \bm \rho} \times \bm K\right)_z = 2 {\,\text{Re\,}} \chi\, E_\alpha  E^*_\beta \frac{\partial^2 h}{\partial\rho_\alpha \partial\rho_\beta} ,
\end{align}
acting per unit area of the membrane. This corresponds to the action of the effective tension $\sigma_{\text{om},\alpha\beta} =  {\,\text{Re\,}} \chi\, (E^*_\alpha  E_\beta + E_\alpha  E^*_\beta) $. Finally, we express the polarizability via the reflection coefficient $r$ as $\chi = r c/[2\pi\rmi \omega (1+r)]$, link the field $\bm E$ to the field of the incident wave $\bm E_0$ by $\bm E = (1+r) \bm E_0$ and obtain
\begin{align}
\sigma_{\text{om},\alpha\beta} =   \frac{c}{\pi\omega}\,\text{Re\,}(E^*_{0,\alpha}  E_{0,\beta}) \,\text{Im\,} r \,
\end{align}
reproducing the first term of Eq.~\eqref{eq:sigma}. 
 
To reproduce the other contribution to the optomechanical tension, we evaluate the Amp\`ere force acting on the membrane. Due to the membrane flexion, the incident electric field, acting on the membrane, changes as
\begin{align}
\bm E_0(h) = \bm E_0 +\rmi \frac{\omega}{c} h \bm E_0 \,.
\end{align}
Suppose now the membrane flexion has the form of harmonic modulation $h(\bm \rho) = h_0 \e^{\rmi \bm q \cdot \bm\rho} + \text{c.c.}$. The induced current then reads, linearly in $\chi$,
\begin{align}
j = -\rmi\omega\chi \bm E_0 + \frac{\omega^2}{c} \chi \bm E_0  ( h_0 \e^{\rmi \bm q \cdot \bm\rho} + \text{c.c.})\,.
\end{align}
To account for radiative corrections, the bare polarizability $\chi$ has to be replaced by the dressed one, which is easily expressed via the light reflection coefficients. Since the current has an in-plane wave vector, the reflection coefficient for oblique incidence should be used,
\begin{align}
\bm j = -\frac{rc}{2\pi} \bm E_0 - \frac{\rmi\omega}{2\pi} \left[ \frac{r_p(\theta)}{\cos\theta} \bm E_{0,\parallel}   +  r_s(\theta) \cos\theta \bm E_{0,\perp} \right] ( h_0 \e^{\rmi \bm q \cdot \bm\rho} + \text{c.c.}) \,,
\end{align}
where $ \bm E_{0,\parallel}$ and $\bm E_{0,\perp}$ are the components of $\bm E_0$ parallel and perpendicular to $\bm q$. 
The Amp\`ere force acting on the current is given by $f_z  =(2/c) \text{Re\,} [\bm j \times \bm B(h)^*]_z$, where
\begin{align}
\bm B(h) = \bm B_0 + \rmi \frac{\omega}{c} h (1+r) \bm B_0
\end{align}
is the magnetic field at the layer. Then, up to the terms linear in $h$, we get
\begin{align}
f_z  = -\frac{|E_0|^2}{\pi} \,\text{Re\,}r + \frac{\omega}{\pi c} \,\text{Im\,} \left[ \frac{r_p(\theta)}{\cos\theta} |E_{0,\parallel}|^2   +  r_s(\theta) \cos\theta |E_{0,\perp}|^2 - r|E_0|^2 \right] ( h_0 \e^{\rmi \bm q \cdot \bm\rho} + \text{c.c.}) \,.
\end{align}
While the first term describes the homogeneous light pressure, the second, being proportional to $h$, is a contribution to the optomechanical back-action force.  Next, we suppose that $\theta \ll 1$, use the expansion
\begin{align}
&r_s(\theta) \cos\theta= \frac{r}{1+r(1-1/\cos\theta)} = r + r^2\frac{ \theta^2}2 + \ldots\\
&\frac{r_p(\theta)}{\cos\theta} = \frac{r}{1+r(1 -\cos\theta)} = r -r^2 \frac{\theta^2}2 + \ldots
\end{align}
and obtain the quadratic in $q$ contribution
\begin{align}
f_z  =- \frac{c q^2}{2\pi \omega} \,\text{Im\,}(r^2) \left(|E_{0,\parallel}|^2  - |E_{0,\perp}|^2 \right) ( h_0 \e^{\rmi \bm q \cdot \bm\rho} + \text{c.c.}) \,,
\end{align}
which reproduces the second term of  Eq.~\eqref{eq:sigma}.

\section{S3. Calculation of the optomechanical tension from the Maxwell stress tensor}
  Yet another approach to calculate the optomechanical tension is provided by the momentum conservation law in classical electrodynamics.
  Namely, we solve the Maxwell equations directly  accounting the membrane flection $h$ as a perturbation~\cite{KerkerPRX}. Next, we  use the Maxwell stress tensor to determine the linear-in-$h$ force acting on the membrane. The whole procedure is rather straightforward but requires accurate handling of 
 the multiple discontinuities and singularities of  the electromagnetic field components at the membrane surface.
 The details are given below.

\subsection{Light scattering on the  deformed membrane}
\label{subsec:permittivity_deformed}

Our aim is to solve the wave equation for the electric field
\begin{equation}\label{eq:main}
\rot\rot\bm E=-\frac1{c^{2}}\frac{\partial^{2}\bm D}{\partial t^{2}}\:.
\end{equation}
The general  geometry of the problem is shown in Fig.~\ref{fig:S:coors}.
The vibration of the membrane surface 
$$
z = h(x, y)\, \e^{-\rmi\Omega t} + \text{c.c.}
$$
with the frequency $\Omega$ modifies the spatial distribution of the permittivity tensor $\eps(\bm r)$, and thus affects the displacement
$\bm D=\eps \bm E$, leading to the appearance of Stokes and anti-Stokes scattered waves,
oscillating at the frequencies $\omega_\text{S}=\omega-\Omega$, $\omega_\text{aS}=\omega+\Omega$.

We choose the $x$ axis along the phonon wave vector $\bm q$. 
Since the calculation of the optical force is rather lengthy, below we reproduce only the result for the normal incidence ($\theta=0$). 

In order to derive the explicit expression for the permittivity tensor of the deformed membrane, we introduce the 
 the normals $\bm{n}(x,y)$ to the membrane surface 
\begin{equation}
	\label{eq:h_n}
	\bm{n}(x, y)\equiv
	\begin{pmatrix}
		n_x \\
		n_y \\
		n_z
	\end{pmatrix}
 = \frac{1}{\sqrt{1+(\partial_x h)^2 + (\partial_y h)^2}}
	\begin{pmatrix}
		- \partial_x h \\
		- \partial_y h \\
		1
	\end{pmatrix}\:.	
\end{equation}
The permittivity tensor is then compactly expressed as
\begin{equation}
	\varepsilon_{ij}(\bm r) = \delta_{ij} + 4\pi\chi \delta(z - h) \, (\delta_{ij} - n_i n_j)\:, \quad i,j=x,y,z\:.
\end{equation}
where $\chi$ is the in-plane susceptibility of the membrane while the out-of-plane susceptibility is supposed zero. The projection operator  $\delta_{ij} - n_i n_j$ takes into account that the considered thin membrane has the electric dipole polarization only along its curved surface.
Using the explicit form of $h = h_0 \e^{\rmi qx-\rmi \Omega t}+\rm c.c.$ we expand this operator up to the linear-in-$h$ terms,
\begin{equation}
	\delta_{ij} - n_i n_j = 
	\begin{pmatrix}
		1 & 0 & \rmi q h_0 \\
		0 & 1 & 0 \\
		\rmi q h_0 & 0 & 0
	\end{pmatrix} \, \e^{\rmi qx-\rmi \Omega t}+\rm c.c.
\end{equation}
Finally, expanding $\delta(z-h) = \delta'(z) - h \delta(z) $, we derive the permittivity tensor,
\begin{equation}
	\label{eq:epsilon01}
	\varepsilon^{(0)}_{\omega} = 
	1
	+
	4\pi\chi \delta(z)
	\begin{pmatrix}
		1& 0& 0\\
		0& 1& 0 \\
		0& 0& 0
	\end{pmatrix}
	\:,\quad
	\varepsilon^{(1)}_{\omega\pm\Omega}
	=
	4\pi\chi
	h_0 \e^{\pm (\rmi q x - \rmi \Omega t)}
	\left[
	\pm \rmi  q \delta(z)
	\begin{pmatrix}
		0& 0& 1 \\
		0& 0& 0 \\
		1& 0& 0
	\end{pmatrix} 
	- \delta'(z)
	\begin{pmatrix}
		1& 0& 0\\
		0& 1& 0 \\
		0& 0& 0
	\end{pmatrix}
	\right]
	\:.
\end{equation}
The linear-in-$h$ terms in the displacement vector $\bm D_\text{aS(S)}$ 
have the form
$
	\bm D^{(1)}_\text{aS(S)} = \varepsilon^{(0)}_{\omega\pm\Omega} \bm E^{(1)}_\text{aS(S)} + \varepsilon^{(1)}_{\omega\pm\Omega}\bm E^{(0)} 
$.
Next, we substitute the displacement vector $\bm D=\bm D^{(0)}\e^{-\rmi\omega t}+\bm D^{(1)}_{\rm S}\e^{-\rmi\omega_{\rm S} t}+\bm D^{(1)}_{\rm aS}\e^{-\rmi\omega_{\rm aS} t}$ into the wave equation Eq.~\eqref{eq:main} and solve for the electric field.

\subsection{Explicit results for the scattered electric fields }
In this section we present the explicit form of the electric field, scattered on the deformed membrane, in the zeroth and first order in the membrane deformation $h$. The field can be obtained from the solution of the Maxwell equations in the previous section. That yields the same results as in \cite{KerkerPRX} but neglects the possible difference between $\chi(\omega)$ and $\chi(\omega)_\text{aS(S)}$. The latter however is not important for the optomechanical tension force that is defined for $\Omega=0$. 

We introduce the basis vectors for electric field of $s$- and $p$-polarized waves 
\begin{equation}
	\label{eq:orts0}
	\bm e_s =\bm e_{y}
	\:,\quad
	\bm e_p = -\bm e_{x}
	\:,\quad
	\bm e_k = \bm e_{x}
	\:,
\end{equation}
for the incident wave and 
\begin{equation}
	\label{eq:orts1}
	\bm e_{s,\text{aS(S)}} = \mp \bm e_{y}
	\:,\quad
	\bm e_{p,\text{aS(S)}} = \pm\cos\theta_\text{aS(S)} \bm e_{x} + \sin\theta_\text{aS(S)}\bm e_{z}
	\:,\quad
	\bm e_{k,\text{aS(S)}} =\mp\sin \theta_\text{aS(S)}\bm e_{x}+ \cos \theta_\text{aS(S)} \bm e_{z}
	\:.
\end{equation}
for the anti-Stokes (aS) and Stokes (S) scattered waves. The incident wave is a superposition 
of two light polarizations
\begin{equation}
	\label{eq:Ei}
	\bm E =E_{0}( \bm{e}_s \sin\varphi + \bm{e}_p \cos\varphi) \:.
\end{equation}
The electric field, describing light scattering from the flat undeformed membrane is
\begin{align}
	\label{eq:Es0}
	\bm{E}^{(0)} &= \left( \bm{e}_s \e^{\rmi \omega \bm{e}_k \bm{r}  - \rmi \omega t}  +  r \bm{e}_s \e^{\rmi \omega \bm{e}_k \hat{Z} \bm{r}  - \rmi \omega t} \right) E_0 \sin\varphi
\\	
	&+\left( \bm{e}_p \e^{\rmi \omega \bm{e}_k \bm{r}  - \rmi \omega t}  +  r \hat{Z} \bm{e}_p \e^{\rmi \omega \bm{e}_k \hat{Z} \bm{r}  - \rmi \omega t} \right)E_0 \cos\varphi  \nonumber\:.
\end{align}
Here, we have introduced an auxiliary operator  $\hat{Z}\equiv {\rm diag}\:(1,1,\sign z)$ accounting  for the altering sign along $z$ and set $c=1$ for simplicity. 
The linear-in-$h$ anti-Stokes (Stokes) scattering can be described by the Jones matrix~\cite{KerkerPRX}
\begin{align}\label{eq:huyg}
&S_{ss}(z) = \rmi \frac{\omega_\text{aS(S)}}{c}h_{0}\, \cos\theta \cos\phi_\text{aS(S)} \left[ r_s(\theta_\text{aS(S)},\omega_\text{aS(S)}) - \sign(z) r_s(\theta,\omega)  \right] ,\\\nonumber
&S_{ps}(z) = \rmi \frac{\omega_\text{aS(S)}}{c}h_{0}\, \frac{\cos\theta \sin\phi_\text{aS(S)}}{\cos\theta_\text{aS(S)}} \left[ r_p(\theta_\text{aS(S)},\omega_\text{aS(S)}) -\sign(z) r_s(\theta,\omega)  \right] ,\\\nonumber
&S_{sp}(z) = -\rmi \frac{\omega_\text{aS(S)}}{c}h_{0}\, \sin\phi_\text{aS(S)} \left[ r_s(\theta_\text{aS(S)},\omega_\text{aS(S)}) - \sign(z) r_p(\theta,\omega)  \right] ,\\\nonumber
&S_{pp}(z) = \rmi \frac{\omega_\text{aS(S)}}{c}h_{0}\, \frac{\cos\phi_\text{aS(S)} -\sin\theta \sin\theta_\text{aS(S)}}{\cos\theta_\text{aS(S)}} \left[ r_p(\theta_\text{aS(S)},\omega_\text{aS(S)}) - \sign(z) r_p(\theta,\omega) \right] ,
\end{align}
where $\theta_\text{aS(S)}$ is the angle between $\bm k_\text{aS(S)}$ and $\bm z$, $\phi_{\text{aS(S)}}$ is the angle between $\bm{k}_{\text{aS(S)}\perp}$ and $\bm q$, see Fig.~\ref{fig:S:coors}, and 
\[
r_{s,p}(\omega,\theta) = \frac{2\pi\rmi\omega\chi(\omega)}{c \cos^{\pm1}\theta-2\pi\rmi\omega\chi(\omega)}\]
 are the reflection coefficients for the two polarizations.
In case of normal incidence $r_p=r_s=r$ and $\phi_\text{aS(S)} = 0$.

Then, the Jones matrix is diagonal and the electric field of the anti-Stokes (Stokes) scattered light assumes the form
 \begin{align}
	\label{eq:Es1}
	\bm{E}_{\text{aS(S)}}^{(1)} &=  S_{ss,\text{aS(S)}}(z) \bm{e}_{s,\text{aS(S)}} \e^{\rmi \omega_{(a)S} \bm{e}_{k,\text{aS(S)}} \hat{Z} \bm{r}  - \rmi \omega_{\text{aS(S)}} t}  E_0 \sin\varphi
	\\
	&+
	 S_{pp,\text{aS(S)}}(z) \hat{Z}\bm{e}_{p,\text{aS(S)}} \e^{\rmi \omega_{\text{aS(S)}} \bm{e}_{k,\text{aS(S)}} \hat{Z} \bm{r}  - \rmi \omega_{\text{aS(S)}} t} 
	 E_0 \cos\varphi
	 \:,\nonumber
\end{align}
where we introduced the operator 
\[
 \hat{Z} = 
 \begin{pmatrix}
 	1 & 0 & 0 \\
	0 & 1 & 0 \\
	0 & 0 & \sign(z)
 \end{pmatrix}
\]
to account for the alternating sign of the $z$ component of $\bm{e}_{p,\text{aS(S)}\hookleftarrow}$ and $\bm{e}_{p,\text{aS(S)}\rightarrow}$, see Fig.~\ref{fig:S:coors}.

The explicit form of the magnetic field is readily evaluated from Eqs.~\eqref{eq:Es0} and~\eqref{eq:Es1}.

\subsection{Derivation of the optical force}
 Once the 
   electromagnetic field around the trembling membrane is found, we proceed to the calculation of the optical forces. We start with evaluation of the momentum density of the electromagnetic field
 \begin{align}
 \bm S = \frac{1}{4 \pi c}\bm E \times \bm H
 \end{align}
and the Maxwell stress tensor of the form~\cite{Pfeifer07}
 \begin{equation}
	\hat{\sigma} =\frac1{4 \pi } \left[ \bm E \otimes \bm D + \bm H \otimes \bm H - \frac12 ( \bm E \cdot \bm D  + \bm H \cdot \bm H) \right] 
\end{equation}
that determines the flux of the field momentum. 
The force acting  on the matter in a small volume $dV$ is expressed as
\begin{equation}
	\label{eq:mcl} 
	\rmd F_i = \rmd V \left( \partial_x \sigma_{xi} + \partial_y \sigma_{yi} + \partial_z \sigma_{zi} - \frac{\rmd S_i}{\rmd t} \right) \:,~ i = x, y, z \:.
\end{equation}
Then, the $z$-component of the surface force density $f_z$,  which describes the back-action for flexural vibrations, reads
\begin{equation}
	\label{eq:fz}
	f_z (x, y) 
	= \int\limits_{-0}^{+0} \rmd z\, \frac{\rmd F_z}{\rmd V}  \:.
\end{equation}

The electromagnetic fields $\bm E,\bm D,\bm H$, the vibration amplitude $h$ and the components of the stress-energy tensor are real. Thus, we need to calculate only, e.g., the anti-Stokes component of the optical force $f_{\rm aS} = f_{\Omega} \propto \exp(-\rmi \Omega t)$ to obtain the linear-in-$h$ contribution to the optical force $f^{(1)} = f_{\Omega} + f_{-\Omega}^*$ acting on the membrane.
As the force $f$ is the sum of the products $\sum_{ij} \int \rmd z  F_i F_j$, where $\bm F = \bm  F^{(0)} +\bm  F^{(1)}_{\rm aS} +\bm  F^{(1)}_{\rm S} + \rm c.c. $, its anti-Stokes component assumes the general form
\begin{equation}
	f_{\rm aS}^{(1)} = \sum_{ij} \int \rmd z (F_i F_j)^{(1)}_{\rm aS} = \sum_{ij} \int \rmd z ( F_i^{(0)} {F_{j,\rm S}^{(1)}}^* + F_{i,\rm aS}^{(1)} {F_j^{(0)}}^* + {F_i^{(0)}}^* F_{j,\rm aS}^{(1)} + {F_{i,S}^{(1)}}^* F_j^{(0)}) \propto e^{-\rmi \Omega t} \:,
\end{equation}
where $F_{i}$ stands for the Cartesian component of any of the fields $\bm E,\bm D$ or $\bm H$.

For normal light incidence and phonon propagating along $x$, the second term in~Eq.~\eqref{eq:mcl} vanishes due to the uniformity along $y$ ($\partial_y \to 0$). 
There exist 10 products to be evaluated in total.
Nonzero contributions to the optical force $f_z$ come only from singular terms $\propto \delta(z)$ in $\sigma_{xz}$ or $S_z$: $\{E_xD_z, H_xH_z, \quad E_xH_y, E_yH_x\}, $ and discontinuous terms $\propto \sign(z)$ in $\sigma_{zz}$: $\{E_zD_z, E_xD_x, E_yD_y, \quad H_zH_z, H_xH_x, H_yH_y\}$, that we present below.

Singular terms in $E_xD_z$ and $H_xH_z$ (components of $\sigma_{xz}$):
\[
	(E_x D_z)_{\rm aS}^{(1)} =  4\pi \rmi E_0^2 (\chi+\chi^*) q h_0 \e^{\rmi q x - \rmi \Omega t} \left| 1+r \right|^2 \cos^2\varphi\, \delta(z) \: .
\]

Singular terms in $E_x H_y$ and $E_yH_x$ (components of $S_z$):
\begin{gather*}
	(E_x H_y)_{\rm aS}^{(1)} = - 4 E_0^2 h_0 \e^{\rmi q x - \rmi \Omega t} \left(\text{Re}(r) + |r|^2 \right) \cos^2\varphi\, \delta(z) \:,\\
	(E_y H_x)_{\rm aS}^{(1)} = 4 E_0^2 h_0 \e^{\rmi q x - \rmi \Omega t} \left(\text{Re}(r) + |r|^2 \right) \sin^2\varphi\, \delta(z) \:.
\end{gather*}

Discontinuous terms in $E_iD_i$ and $H_iH_i$ (components of $\sigma_{zz}$):
\begin{gather*}
	(E_x D_x)_{\rm aS}^{(1)} = \frac{2 \rmi E_0^2}{c} h_0 \e^{ \rmi q x - \rmi \Omega t} \left(r^* \omega_\text{S} -r \omega_\text{aS}  + |r|^2 (\omega_\text{S}  - \omega_{\rm aS}) \right) \cos^2\varphi\,  \sign z \:,\\
	(E_y D_y)_{\rm aS}^{(1)} =  \frac{2 \rmi E_0^2}{c} h_0 \e^{ \rmi q x - \rmi \Omega t} \left( r^* \omega_\text{S}  -r \omega_\text{aS}  + |r|^2 (\omega_{\rm S} - \omega_{\rm aS}) \right) \sin^2\varphi\, \sign z \:,\\
	(H_x H_x)_{\rm aS}^{(1)} = -\frac{2\rmi E_0^2}{c} h_0 \e^{\rmi q x - \rmi \Omega t} 
	\left[
		({r_{\rm S}^{s}}^* - |r|^2) \omega_{\rm S} \cos \theta_{\rm S}
		-
		({r_{\rm aS}^s} - |r|^2) \omega_{\rm aS} \cos \theta_{\rm aS}
	\right]
	\sin^2\varphi\, \sign z \:,\\
	(H_y H_y)_{\rm aS}^{(1)} = - \frac{2 \rmi E_0^2}{c} h_0 \e^{\rmi q x - \rmi \Omega t} 
	\left[
		({r_{\rm S}^p}^* - |r|^2) \frac{\omega_{\rm S}} {\cos \theta_{\rm S} }
		-
		({r_{\rm aS}^p} - |r|^2) \frac{ \omega_{\rm aS} }{ \cos\theta_{\rm aS} }
	\right]
	\cos^2\varphi\, \sign z \:, \\
	(H_x H_z)_{\rm aS}^{(1)}=(H_z H_z)_{\rm aS}^{(1)}=(E_z D_z)_{\rm aS}^{(1)} =0 \:.
\end{gather*}

After the integration over $\rmd z$ we finally derive the optical force
\begin{multline}
    f_z(x,y) = 
	 \frac{E_0^2}{4 \pi c}
    h_0 \e^{ \rmi q x - \rmi \Omega t}
    \left\{
    - 4 \pi c (\chi+\chi^*) q^2 \left| 1+r \right|^2 \cos^2\varphi
    - 2 \rmi \left( r^* \omega_{\rm S} -r \omega_{\rm aS} + |r|^2 (\omega_{\rm S} - \omega_{\rm aS})  \right) +
    \right.
    \\
    +2 \rmi 
    \left[
        ({r_{\rm S}^{s}}^* - |r|^2) \omega_{\rm S} \cos \theta_{\rm S}
        -
        ({r_{\rm aS}^s} - |r|^2) \omega_{\rm aS} \cos \theta_{\rm aS}
    \right]
    \sin^2\varphi
    + 2 \rmi  
    \left[
        ({r_{\rm S}^p}^* - |r|^2) \frac{\omega_{\rm S}} {\cos \theta_{\rm S}}
        -
        ({r_{\rm aS}^p} - |r|^2) \frac{ \omega_{\rm aS} }{ \cos\theta_{\rm aS} }
    \right]
    \cos^2\varphi 
    \\
    \left.
	- 4 \rmi\Omega \left(\text{Re}\,r + |r|^2 \right)
    \right\}
    \:.
\end{multline}
Substituting
$$
	\chi = \frac{\rmi c}{2\pi\omega} \frac{r}{1+r}
$$ 
we obtain 
\begin{align}
 f_z(x, y) = -\Sigma_\text{om} h_0 \e^{\rmi q x - \rmi \Omega t}
 \end{align}
 where $\Sigma_\text{om}$ matches the one determined by Eq.~\eqref{eq:S} of the main text that was calculated using the diagrammatic approach of Section S1. 
 Specifically, in the static limit of vanishing phonon frequency, $\Omega\to 0$, when $\omega_{\rm aS}=\omega_{\rm S}\equiv \omega, \theta_{\rm aS}=\theta_{\rm S}$, we recover Eq.~\eqref{eq:sigma} from the main text.


\begin{thebibliography}{44}%
\makeatletter
\providecommand \@ifxundefined [1]{%
 \@ifx{#1\undefined}
}%
\providecommand \@ifnum [1]{%
 \ifnum #1\expandafter \@firstoftwo
 \else \expandafter \@secondoftwo
 \fi
}%
\providecommand \@ifx [1]{%
 \ifx #1\expandafter \@firstoftwo
 \else \expandafter \@secondoftwo
 \fi
}%
\providecommand \natexlab [1]{#1}%
\providecommand \enquote  [1]{``#1''}%
\providecommand \bibnamefont  [1]{#1}%
\providecommand \bibfnamefont [1]{#1}%
\providecommand \citenamefont [1]{#1}%
\providecommand \href@noop [0]{\@secondoftwo}%
\providecommand \href [0]{\begingroup \@sanitize@url \@href}%
\providecommand \@href[1]{\@@startlink{#1}\@@href}%
\providecommand \@@href[1]{\endgroup#1\@@endlink}%
\providecommand \@sanitize@url [0]{\catcode `\\12\catcode `\$12\catcode
  `\&12\catcode `\#12\catcode `\^12\catcode `\_12\catcode `\%12\relax}%
\providecommand \@@startlink[1]{}%
\providecommand \@@endlink[0]{}%
\providecommand \url  [0]{\begingroup\@sanitize@url \@url }%
\providecommand \@url [1]{\endgroup\@href {#1}{\urlprefix }}%
\providecommand \urlprefix  [0]{URL }%
\providecommand \Eprint [0]{\href }%
\providecommand \doibase [0]{http://dx.doi.org/}%
\providecommand \selectlanguage [0]{\@gobble}%
\providecommand \bibinfo  [0]{\@secondoftwo}%
\providecommand \bibfield  [0]{\@secondoftwo}%
\providecommand \translation [1]{[#1]}%
\providecommand \BibitemOpen [0]{}%
\providecommand \bibitemStop [0]{}%
\providecommand \bibitemNoStop [0]{.\EOS\space}%
\providecommand \EOS [0]{\spacefactor3000\relax}%
\providecommand \BibitemShut  [1]{\csname bibitem#1\endcsname}%
\let\auto@bib@innerbib\@empty
\bibitem [{\citenamefont {Nelson}\ \emph {et~al.}(2004)\citenamefont {Nelson},
  \citenamefont {Piran},\ and\ \citenamefont
  {Weinberg}}]{nelson2004statistical}%
  \BibitemOpen
  \bibfield  {author} {\bibinfo {author} {\bibfnamefont {D.~R.}\ \bibnamefont
  {Nelson}}, \bibinfo {author} {\bibfnamefont {T.}~\bibnamefont {Piran}}, \
  and\ \bibinfo {author} {\bibfnamefont {S.}~\bibnamefont {Weinberg}},\
  }\href@noop {} {\emph {\bibinfo {title} {Statistical mechanics of membranes
  and surfaces}}}\ (\bibinfo  {publisher} {World Scientific},\ \bibinfo {year}
  {2004})\BibitemShut {NoStop}%
\bibitem [{\citenamefont {Nicholl}\ \emph {et~al.}(2015)\citenamefont
  {Nicholl}, \citenamefont {Conley}, \citenamefont {Lavrik}, \citenamefont
  {Vlassiouk}, \citenamefont {Puzyrev}, \citenamefont {Sreenivas},
  \citenamefont {Pantelides},\ and\ \citenamefont {Bolotin}}]{Nicholl2013}%
  \BibitemOpen
  \bibfield  {author} {\bibinfo {author} {\bibfnamefont {R.~J.~T.}\
  \bibnamefont {Nicholl}}, \bibinfo {author} {\bibfnamefont {H.~J.}\
  \bibnamefont {Conley}}, \bibinfo {author} {\bibfnamefont {N.~V.}\
  \bibnamefont {Lavrik}}, \bibinfo {author} {\bibfnamefont {I.}~\bibnamefont
  {Vlassiouk}}, \bibinfo {author} {\bibfnamefont {Y.~S.}\ \bibnamefont
  {Puzyrev}}, \bibinfo {author} {\bibfnamefont {V.~P.}\ \bibnamefont
  {Sreenivas}}, \bibinfo {author} {\bibfnamefont {S.~T.}\ \bibnamefont
  {Pantelides}}, \ and\ \bibinfo {author} {\bibfnamefont {K.~I.}\ \bibnamefont
  {Bolotin}},\ }\bibfield  {title} {\enquote {\bibinfo {title} {The effect of
  intrinsic crumpling on the mechanics of free-standing graphene},}\ }\href
  {https://doi.org/10.1038/ncomms9789} {\bibfield  {journal} {\bibinfo
  {journal} {Nat. Commun.}\ }\textbf {\bibinfo {volume} {6}},\ \bibinfo {pages}
  {8789} (\bibinfo {year} {2015})}\BibitemShut {NoStop}%
\bibitem [{\citenamefont {Liu}\ and\ \citenamefont {Wu}(2016)}]{liu_wu_2016}%
  \BibitemOpen
  \bibfield  {author} {\bibinfo {author} {\bibfnamefont {K.}~\bibnamefont
  {Liu}}\ and\ \bibinfo {author} {\bibfnamefont {J.}~\bibnamefont {Wu}},\
  }\bibfield  {title} {\enquote {\bibinfo {title} {Mechanical properties of
  two-dimensional materials and heterostructures},}\ }\href {\doibase
  10.1557/jmr.2015.324} {\bibfield  {journal} {\bibinfo  {journal} {Journal of
  Materials Research}\ }\textbf {\bibinfo {volume} {31}},\ \bibinfo {pages}
  {832} (\bibinfo {year} {2016})}\BibitemShut {NoStop}%
\bibitem [{\citenamefont {Khestanova}\ \emph {et~al.}(2016)\citenamefont
  {Khestanova}, \citenamefont {Guinea}, \citenamefont {Fumagalli},
  \citenamefont {Geim},\ and\ \citenamefont {Grigorieva}}]{Khestanova2016}%
  \BibitemOpen
  \bibfield  {author} {\bibinfo {author} {\bibfnamefont {E.}~\bibnamefont
  {Khestanova}}, \bibinfo {author} {\bibfnamefont {F.}~\bibnamefont {Guinea}},
  \bibinfo {author} {\bibfnamefont {L.}~\bibnamefont {Fumagalli}}, \bibinfo
  {author} {\bibfnamefont {A.~K.}\ \bibnamefont {Geim}}, \ and\ \bibinfo
  {author} {\bibfnamefont {I.~V.}\ \bibnamefont {Grigorieva}},\ }\bibfield
  {title} {\enquote {\bibinfo {title} {Universal shape and pressure inside
  bubbles appearing in van der {W}aals heterostructures},}\ }\href {\doibase
  10.1038/ncomms12587} {\bibfield  {journal} {\bibinfo  {journal} {Nat.
  Commun.}\ }\textbf {\bibinfo {volume} {7}},\ \bibinfo {pages} {12587}
  (\bibinfo {year} {2016})}\BibitemShut {NoStop}%
\bibitem [{\citenamefont {Doussal}\ and\ \citenamefont
  {Radzihovsky}(2018)}]{LeDoussal2018}%
  \BibitemOpen
  \bibfield  {author} {\bibinfo {author} {\bibfnamefont {P.~L.}\ \bibnamefont
  {Doussal}}\ and\ \bibinfo {author} {\bibfnamefont {L.}~\bibnamefont
  {Radzihovsky}},\ }\bibfield  {title} {\enquote {\bibinfo {title} {Anomalous
  elasticity, fluctuations and disorder in elastic membranes},}\ }\href
  {\doibase 10.1016/j.aop.2017.08.033} {\bibfield  {journal} {\bibinfo
  {journal} {Annals of Physics}\ }\textbf {\bibinfo {volume} {392}},\ \bibinfo
  {pages} {340} (\bibinfo {year} {2018})}\BibitemShut {NoStop}%
\bibitem [{\citenamefont {Balandin}(2011)}]{Balandin2011}%
  \BibitemOpen
  \bibfield  {author} {\bibinfo {author} {\bibfnamefont {A.~A.}\ \bibnamefont
  {Balandin}},\ }\bibfield  {title} {\enquote {\bibinfo {title} {Thermal
  properties of graphene and nanostructured carbon materials},}\ }\href
  {\doibase 10.1038/nmat3064} {\bibfield  {journal} {\bibinfo  {journal} {Nat.
  Mater.}\ }\textbf {\bibinfo {volume} {10}},\ \bibinfo {pages} {569} (\bibinfo
  {year} {2011})}\BibitemShut {NoStop}%
\bibitem [{\citenamefont {Cepellotti}\ and\ \citenamefont
  {Marzari}(2016)}]{Cepelotti2016}%
  \BibitemOpen
  \bibfield  {author} {\bibinfo {author} {\bibfnamefont {A.}~\bibnamefont
  {Cepellotti}}\ and\ \bibinfo {author} {\bibfnamefont {N.}~\bibnamefont
  {Marzari}},\ }\bibfield  {title} {\enquote {\bibinfo {title} {Thermal
  transport in crystals as a kinetic theory of relaxons},}\ }\href {\doibase
  10.1103/PhysRevX.6.041013} {\bibfield  {journal} {\bibinfo  {journal} {Phys.
  Rev. X}\ }\textbf {\bibinfo {volume} {6}},\ \bibinfo {pages} {041013}
  (\bibinfo {year} {2016})}\BibitemShut {NoStop}%
\bibitem [{\citenamefont {Song}\ \emph {et~al.}(2018)\citenamefont {Song},
  \citenamefont {Liu}, \citenamefont {Liu}, \citenamefont {Wu}, \citenamefont
  {Cheng},\ and\ \citenamefont {Kang}}]{Song2018}%
  \BibitemOpen
  \bibfield  {author} {\bibinfo {author} {\bibfnamefont {H.}~\bibnamefont
  {Song}}, \bibinfo {author} {\bibfnamefont {J.}~\bibnamefont {Liu}}, \bibinfo
  {author} {\bibfnamefont {B.}~\bibnamefont {Liu}}, \bibinfo {author}
  {\bibfnamefont {J.}~\bibnamefont {Wu}}, \bibinfo {author} {\bibfnamefont
  {H.-M.}\ \bibnamefont {Cheng}}, \ and\ \bibinfo {author} {\bibfnamefont
  {F.}~\bibnamefont {Kang}},\ }\bibfield  {title} {\enquote {\bibinfo {title}
  {Two-dimensional materials for thermal management applications},}\ }\href
  {\doibase https://doi.org/10.1016/j.joule.2018.01.006} {\bibfield  {journal}
  {\bibinfo  {journal} {Joule}\ }\textbf {\bibinfo {volume} {2}},\ \bibinfo
  {pages} {442} (\bibinfo {year} {2018})}\BibitemShut {NoStop}%
\bibitem [{\citenamefont {Kang}\ \emph {et~al.}(2018)\citenamefont {Kang},
  \citenamefont {Li}, \citenamefont {Wu}, \citenamefont {Nguyen},\ and\
  \citenamefont {Hu}}]{Kang2018}%
  \BibitemOpen
  \bibfield  {author} {\bibinfo {author} {\bibfnamefont {J.~S.}\ \bibnamefont
  {Kang}}, \bibinfo {author} {\bibfnamefont {M.}~\bibnamefont {Li}}, \bibinfo
  {author} {\bibfnamefont {H.}~\bibnamefont {Wu}}, \bibinfo {author}
  {\bibfnamefont {H.}~\bibnamefont {Nguyen}}, \ and\ \bibinfo {author}
  {\bibfnamefont {Y.}~\bibnamefont {Hu}},\ }\bibfield  {title} {\enquote
  {\bibinfo {title} {Experimental observation of high thermal conductivity in
  boron arsenide},}\ }\href {\doibase 10.1126/science.aat5522} {\bibfield
  {journal} {\bibinfo  {journal} {Science}\ }\textbf {\bibinfo {volume}
  {361}},\ \bibinfo {pages} {575} (\bibinfo {year} {2018})}\BibitemShut
  {NoStop}%
\bibitem [{\citenamefont {Seol}\ \emph {et~al.}(2010)\citenamefont {Seol},
  \citenamefont {Jo}, \citenamefont {Moore}, \citenamefont {Lindsay},
  \citenamefont {Aitken}, \citenamefont {Pettes}, \citenamefont {Li},
  \citenamefont {Yao}, \citenamefont {Huang}, \citenamefont {Broido},
  \citenamefont {Mingo}, \citenamefont {Ruoff},\ and\ \citenamefont
  {Shi}}]{Seol2010}%
  \BibitemOpen
  \bibfield  {author} {\bibinfo {author} {\bibfnamefont {J.~H.}\ \bibnamefont
  {Seol}}, \bibinfo {author} {\bibfnamefont {I.}~\bibnamefont {Jo}}, \bibinfo
  {author} {\bibfnamefont {A.~L.}\ \bibnamefont {Moore}}, \bibinfo {author}
  {\bibfnamefont {L.}~\bibnamefont {Lindsay}}, \bibinfo {author} {\bibfnamefont
  {Z.~H.}\ \bibnamefont {Aitken}}, \bibinfo {author} {\bibfnamefont {M.~T.}\
  \bibnamefont {Pettes}}, \bibinfo {author} {\bibfnamefont {X.}~\bibnamefont
  {Li}}, \bibinfo {author} {\bibfnamefont {Z.}~\bibnamefont {Yao}}, \bibinfo
  {author} {\bibfnamefont {R.}~\bibnamefont {Huang}}, \bibinfo {author}
  {\bibfnamefont {D.}~\bibnamefont {Broido}}, \bibinfo {author} {\bibfnamefont
  {N.}~\bibnamefont {Mingo}}, \bibinfo {author} {\bibfnamefont {R.~S.}\
  \bibnamefont {Ruoff}}, \ and\ \bibinfo {author} {\bibfnamefont
  {L.}~\bibnamefont {Shi}},\ }\bibfield  {title} {\enquote {\bibinfo {title}
  {Two-dimensional phonon transport in supported graphene},}\ }\href {\doibase
  10.1126/science.1184014} {\bibfield  {journal} {\bibinfo  {journal}
  {Science}\ }\textbf {\bibinfo {volume} {328}},\ \bibinfo {pages} {213}
  (\bibinfo {year} {2010})}\BibitemShut {NoStop}%
\bibitem [{\citenamefont {Ackerman}\ \emph {et~al.}(2016)\citenamefont
  {Ackerman}, \citenamefont {Kumar}, \citenamefont {Neek-Amal}, \citenamefont
  {Thibado}, \citenamefont {Peeters},\ and\ \citenamefont
  {Singh}}]{Ackerman2016}%
  \BibitemOpen
  \bibfield  {author} {\bibinfo {author} {\bibfnamefont {M.~L.}\ \bibnamefont
  {Ackerman}}, \bibinfo {author} {\bibfnamefont {P.}~\bibnamefont {Kumar}},
  \bibinfo {author} {\bibfnamefont {M.}~\bibnamefont {Neek-Amal}}, \bibinfo
  {author} {\bibfnamefont {P.~M.}\ \bibnamefont {Thibado}}, \bibinfo {author}
  {\bibfnamefont {F.~M.}\ \bibnamefont {Peeters}}, \ and\ \bibinfo {author}
  {\bibfnamefont {S.}~\bibnamefont {Singh}},\ }\bibfield  {title} {\enquote
  {\bibinfo {title} {Anomalous dynamical behavior of freestanding graphene
  membranes},}\ }\href {\doibase 10.1103/PhysRevLett.117.126801} {\bibfield
  {journal} {\bibinfo  {journal} {Phys. Rev. Lett.}\ }\textbf {\bibinfo
  {volume} {117}},\ \bibinfo {pages} {126801} (\bibinfo {year}
  {2016})}\BibitemShut {NoStop}%
\bibitem [{\citenamefont {Gornyi}\ \emph {et~al.}(2015)\citenamefont {Gornyi},
  \citenamefont {Kachorovskii},\ and\ \citenamefont {Mirlin}}]{Gornyi2015}%
  \BibitemOpen
  \bibfield  {author} {\bibinfo {author} {\bibfnamefont {I.~V.}\ \bibnamefont
  {Gornyi}}, \bibinfo {author} {\bibfnamefont {V.~Y.}\ \bibnamefont
  {Kachorovskii}}, \ and\ \bibinfo {author} {\bibfnamefont {A.~D.}\
  \bibnamefont {Mirlin}},\ }\bibfield  {title} {\enquote {\bibinfo {title}
  {Rippling and crumpling in disordered free-standing graphene},}\ }\href
  {\doibase 10.1103/PhysRevB.92.155428} {\bibfield  {journal} {\bibinfo
  {journal} {Phys. Rev. B}\ }\textbf {\bibinfo {volume} {92}},\ \bibinfo
  {pages} {155428} (\bibinfo {year} {2015})}\BibitemShut {NoStop}%
\bibitem [{\citenamefont {Lee}\ \emph {et~al.}(2015)\citenamefont {Lee},
  \citenamefont {Broido}, \citenamefont {Esfarjani},\ and\ \citenamefont
  {Chen}}]{Broido2015}%
  \BibitemOpen
  \bibfield  {author} {\bibinfo {author} {\bibfnamefont {S.}~\bibnamefont
  {Lee}}, \bibinfo {author} {\bibfnamefont {D.}~\bibnamefont {Broido}},
  \bibinfo {author} {\bibfnamefont {K.}~\bibnamefont {Esfarjani}}, \ and\
  \bibinfo {author} {\bibfnamefont {G.}~\bibnamefont {Chen}},\ }\bibfield
  {title} {\enquote {\bibinfo {title} {Hydrodynamic phonon transport in
  suspended graphene},}\ }\href {\doibase 10.1038/ncomms7290} {\bibfield
  {journal} {\bibinfo  {journal} {Nat. Commun.}\ }\textbf {\bibinfo {volume}
  {6}},\ \bibinfo {pages} {6290} (\bibinfo {year} {2015})}\BibitemShut
  {NoStop}%
\bibitem [{\citenamefont {Burmistrov}\ \emph
  {et~al.}(2018{\natexlab{a}})\citenamefont {Burmistrov}, \citenamefont
  {Kachorovskii}, \citenamefont {Gornyi},\ and\ \citenamefont
  {Mirlin}}]{Burmistrov2018}%
  \BibitemOpen
  \bibfield  {author} {\bibinfo {author} {\bibfnamefont {I.}~\bibnamefont
  {Burmistrov}}, \bibinfo {author} {\bibfnamefont {V.~Y.}\ \bibnamefont
  {Kachorovskii}}, \bibinfo {author} {\bibfnamefont {I.}~\bibnamefont
  {Gornyi}}, \ and\ \bibinfo {author} {\bibfnamefont {A.}~\bibnamefont
  {Mirlin}},\ }\bibfield  {title} {\enquote {\bibinfo {title} {Differential
  {P}oisson's ratio of a crystalline two-dimensional membrane},}\ }\href
  {\doibase https://doi.org/10.1016/j.aop.2018.07.009} {\bibfield  {journal}
  {\bibinfo  {journal} {Annals of Physics}\ }\textbf {\bibinfo {volume}
  {396}},\ \bibinfo {pages} {119} (\bibinfo {year}
  {2018}{\natexlab{a}})}\BibitemShut {NoStop}%
\bibitem [{\citenamefont {Burmistrov}\ \emph
  {et~al.}(2018{\natexlab{b}})\citenamefont {Burmistrov}, \citenamefont
  {Gornyi}, \citenamefont {Kachorovskii}, \citenamefont {Katsnelson},
  \citenamefont {Los},\ and\ \citenamefont {Mirlin}}]{Kachorovskii2018}%
  \BibitemOpen
  \bibfield  {author} {\bibinfo {author} {\bibfnamefont {I.~S.}\ \bibnamefont
  {Burmistrov}}, \bibinfo {author} {\bibfnamefont {I.~V.}\ \bibnamefont
  {Gornyi}}, \bibinfo {author} {\bibfnamefont {V.~Y.}\ \bibnamefont
  {Kachorovskii}}, \bibinfo {author} {\bibfnamefont {M.~I.}\ \bibnamefont
  {Katsnelson}}, \bibinfo {author} {\bibfnamefont {J.~H.}\ \bibnamefont {Los}},
  \ and\ \bibinfo {author} {\bibfnamefont {A.~D.}\ \bibnamefont {Mirlin}},\
  }\bibfield  {title} {\enquote {\bibinfo {title} {Stress-controlled {P}oisson
  ratio of a crystalline membrane: Application to graphene},}\ }\href {\doibase
  10.1103/PhysRevB.97.125402} {\bibfield  {journal} {\bibinfo  {journal} {Phys.
  Rev. B}\ }\textbf {\bibinfo {volume} {97}},\ \bibinfo {pages} {125402}
  (\bibinfo {year} {2018}{\natexlab{b}})}\BibitemShut {NoStop}%
\bibitem [{\citenamefont {Wang}\ \emph {et~al.}(2018)\citenamefont {Wang},
  \citenamefont {Chernikov}, \citenamefont {Glazov}, \citenamefont {Heinz},
  \citenamefont {Marie}, \citenamefont {Amand},\ and\ \citenamefont
  {Urbaszek}}]{Glazov2018}%
  \BibitemOpen
  \bibfield  {author} {\bibinfo {author} {\bibfnamefont {G.}~\bibnamefont
  {Wang}}, \bibinfo {author} {\bibfnamefont {A.}~\bibnamefont {Chernikov}},
  \bibinfo {author} {\bibfnamefont {M.~M.}\ \bibnamefont {Glazov}}, \bibinfo
  {author} {\bibfnamefont {T.~F.}\ \bibnamefont {Heinz}}, \bibinfo {author}
  {\bibfnamefont {X.}~\bibnamefont {Marie}}, \bibinfo {author} {\bibfnamefont
  {T.}~\bibnamefont {Amand}}, \ and\ \bibinfo {author} {\bibfnamefont
  {B.}~\bibnamefont {Urbaszek}},\ }\bibfield  {title} {\enquote {\bibinfo
  {title} {Colloquium: Excitons in atomically thin transition metal
  dichalcogenides},}\ }\href {\doibase 10.1103/RevModPhys.90.021001} {\bibfield
   {journal} {\bibinfo  {journal} {Rev. Mod. Phys.}\ }\textbf {\bibinfo
  {volume} {90}},\ \bibinfo {pages} {021001} (\bibinfo {year}
  {2018})}\BibitemShut {NoStop}%
\bibitem [{\citenamefont {Bertolazzi}\ \emph {et~al.}(2011)\citenamefont
  {Bertolazzi}, \citenamefont {Brivio},\ and\ \citenamefont
  {Kis}}]{Bertolazzi2011}%
  \BibitemOpen
  \bibfield  {author} {\bibinfo {author} {\bibfnamefont {S.}~\bibnamefont
  {Bertolazzi}}, \bibinfo {author} {\bibfnamefont {J.}~\bibnamefont {Brivio}},
  \ and\ \bibinfo {author} {\bibfnamefont {A.}~\bibnamefont {Kis}},\ }\bibfield
   {title} {\enquote {\bibinfo {title} {Stretching and breaking of ultrathin
  {MoS${}_{2}$}},}\ }\href {\doibase 10.1021/nn203879f} {\bibfield  {journal}
  {\bibinfo  {journal} {ACS Nano}\ }\textbf {\bibinfo {volume} {5}},\ \bibinfo
  {pages} {9703--9709} (\bibinfo {year} {2011})}\BibitemShut {NoStop}%
\bibitem [{\citenamefont {Castellanos-Gomez}\ \emph {et~al.}(2013)\citenamefont
  {Castellanos-Gomez}, \citenamefont {van Leeuwen}, \citenamefont {Buscema},
  \citenamefont {van~der Zant}, \citenamefont {Steele},\ and\ \citenamefont
  {Venstra}}]{Gomez2013}%
  \BibitemOpen
  \bibfield  {author} {\bibinfo {author} {\bibfnamefont {A.}~\bibnamefont
  {Castellanos-Gomez}}, \bibinfo {author} {\bibfnamefont {R.}~\bibnamefont {van
  Leeuwen}}, \bibinfo {author} {\bibfnamefont {M.}~\bibnamefont {Buscema}},
  \bibinfo {author} {\bibfnamefont {H.~S.~J.}\ \bibnamefont {van~der Zant}},
  \bibinfo {author} {\bibfnamefont {G.~A.}\ \bibnamefont {Steele}}, \ and\
  \bibinfo {author} {\bibfnamefont {W.~J.}\ \bibnamefont {Venstra}},\
  }\bibfield  {title} {\enquote {\bibinfo {title} {Single-layer {MoS${}_{2}$}
  mechanical resonators},}\ }\href {\doibase 10.1002/adma.201303569} {\bibfield
   {journal} {\bibinfo  {journal} {Advanced Materials}\ }\textbf {\bibinfo
  {volume} {25}},\ \bibinfo {pages} {6719} (\bibinfo {year}
  {2013})}\BibitemShut {NoStop}%
\bibitem [{\citenamefont {Inoue}\ \emph {et~al.}(2017)\citenamefont {Inoue},
  \citenamefont {Anno}, \citenamefont {Imakita}, \citenamefont {Takei},
  \citenamefont {Arie},\ and\ \citenamefont {Akita}}]{Akita2017}%
  \BibitemOpen
  \bibfield  {author} {\bibinfo {author} {\bibfnamefont {T.}~\bibnamefont
  {Inoue}}, \bibinfo {author} {\bibfnamefont {Y.}~\bibnamefont {Anno}},
  \bibinfo {author} {\bibfnamefont {Y.}~\bibnamefont {Imakita}}, \bibinfo
  {author} {\bibfnamefont {K.}~\bibnamefont {Takei}}, \bibinfo {author}
  {\bibfnamefont {T.}~\bibnamefont {Arie}}, \ and\ \bibinfo {author}
  {\bibfnamefont {S.}~\bibnamefont {Akita}},\ }\bibfield  {title} {\enquote
  {\bibinfo {title} {Resonance control of a graphene drum resonator in a
  nonlinear regime by a standing wave of light},}\ }\href {\doibase
  10.1021/acsomega.7b00699} {\bibfield  {journal} {\bibinfo  {journal} {ACS
  Omega}\ }\textbf {\bibinfo {volume} {2}},\ \bibinfo {pages} {5792} (\bibinfo
  {year} {2017})}\BibitemShut {NoStop}%
\bibitem [{\citenamefont {Bar-Ziv}\ and\ \citenamefont
  {Moses}(1994)}]{BarZiv1994}%
  \BibitemOpen
  \bibfield  {author} {\bibinfo {author} {\bibfnamefont {R.}~\bibnamefont
  {Bar-Ziv}}\ and\ \bibinfo {author} {\bibfnamefont {E.}~\bibnamefont
  {Moses}},\ }\bibfield  {title} {\enquote {\bibinfo {title} {Instability and
  ``pearling" states produced in tubular membranes by competition of curvature
  and tension},}\ }\href {\doibase 10.1103/PhysRevLett.73.1392} {\bibfield
  {journal} {\bibinfo  {journal} {Phys. Rev. Lett.}\ }\textbf {\bibinfo
  {volume} {73}},\ \bibinfo {pages} {1392} (\bibinfo {year}
  {1994})}\BibitemShut {NoStop}%
\bibitem [{\citenamefont {Bar-Ziv}\ \emph {et~al.}(1998)\citenamefont
  {Bar-Ziv}, \citenamefont {Moses},\ and\ \citenamefont {Nelson}}]{Barziv1998}%
  \BibitemOpen
  \bibfield  {author} {\bibinfo {author} {\bibfnamefont {R.}~\bibnamefont
  {Bar-Ziv}}, \bibinfo {author} {\bibfnamefont {E.}~\bibnamefont {Moses}}, \
  and\ \bibinfo {author} {\bibfnamefont {P.}~\bibnamefont {Nelson}},\
  }\bibfield  {title} {\enquote {\bibinfo {title} {Dynamic excitations in
  membranes induced by optical tweezers},}\ }\href {\doibase
  https://doi.org/10.1016/S0006-3495(98)77515-0} {\bibfield  {journal}
  {\bibinfo  {journal} {Biophysical Journal}\ }\textbf {\bibinfo {volume}
  {75}},\ \bibinfo {pages} {294} (\bibinfo {year} {1998})}\BibitemShut
  {NoStop}%
\bibitem [{\citenamefont {Sipe}\ \emph {et~al.}(1983)\citenamefont {Sipe},
  \citenamefont {Young}, \citenamefont {Preston},\ and\ \citenamefont {van
  Driel}}]{Sipe1983}%
  \BibitemOpen
  \bibfield  {author} {\bibinfo {author} {\bibfnamefont {J.~E.}\ \bibnamefont
  {Sipe}}, \bibinfo {author} {\bibfnamefont {J.~F.}\ \bibnamefont {Young}},
  \bibinfo {author} {\bibfnamefont {J.~S.}\ \bibnamefont {Preston}}, \ and\
  \bibinfo {author} {\bibfnamefont {H.~M.}\ \bibnamefont {van Driel}},\
  }\bibfield  {title} {\enquote {\bibinfo {title} {Laser-induced periodic
  surface structure. i. theory},}\ }\href {\doibase 10.1103/PhysRevB.27.1141}
  {\bibfield  {journal} {\bibinfo  {journal} {Phys. Rev. B}\ }\textbf {\bibinfo
  {volume} {27}},\ \bibinfo {pages} {1141} (\bibinfo {year}
  {1983})}\BibitemShut {NoStop}%
\bibitem [{\citenamefont {Tsibidis}\ \emph {et~al.}(2012)\citenamefont
  {Tsibidis}, \citenamefont {Barberoglou}, \citenamefont {Loukakos},
  \citenamefont {Stratakis},\ and\ \citenamefont {Fotakis}}]{Stratakis2012}%
  \BibitemOpen
  \bibfield  {author} {\bibinfo {author} {\bibfnamefont {G.~D.}\ \bibnamefont
  {Tsibidis}}, \bibinfo {author} {\bibfnamefont {M.}~\bibnamefont
  {Barberoglou}}, \bibinfo {author} {\bibfnamefont {P.~A.}\ \bibnamefont
  {Loukakos}}, \bibinfo {author} {\bibfnamefont {E.}~\bibnamefont {Stratakis}},
  \ and\ \bibinfo {author} {\bibfnamefont {C.}~\bibnamefont {Fotakis}},\
  }\bibfield  {title} {\enquote {\bibinfo {title} {Dynamics of ripple formation
  on silicon surfaces by ultrashort laser pulses in subablation conditions},}\
  }\href {\doibase 10.1103/PhysRevB.86.115316} {\bibfield  {journal} {\bibinfo
  {journal} {Phys. Rev. B}\ }\textbf {\bibinfo {volume} {86}},\ \bibinfo
  {pages} {115316} (\bibinfo {year} {2012})}\BibitemShut {NoStop}%
\bibitem [{\citenamefont {Paradisanos}\ \emph {et~al.}(2014)\citenamefont
  {Paradisanos}, \citenamefont {Kymakis}, \citenamefont {Fotakis},
  \citenamefont {Kioseoglou},\ and\ \citenamefont {Stratakis}}]{Stratakis2014}%
  \BibitemOpen
  \bibfield  {author} {\bibinfo {author} {\bibfnamefont {I.}~\bibnamefont
  {Paradisanos}}, \bibinfo {author} {\bibfnamefont {E.}~\bibnamefont
  {Kymakis}}, \bibinfo {author} {\bibfnamefont {C.}~\bibnamefont {Fotakis}},
  \bibinfo {author} {\bibfnamefont {G.}~\bibnamefont {Kioseoglou}}, \ and\
  \bibinfo {author} {\bibfnamefont {E.}~\bibnamefont {Stratakis}},\ }\bibfield
  {title} {\enquote {\bibinfo {title} {Intense femtosecond photoexcitation of
  bulk and monolayer {MoS${}_{2}$}},}\ }\href {\doibase 10.1063/1.4891679}
  {\bibfield  {journal} {\bibinfo  {journal} {Appl. Phys. Lett.}\ }\textbf
  {\bibinfo {volume} {105}},\ \bibinfo {pages} {041108} (\bibinfo {year}
  {2014})}\BibitemShut {NoStop}%
\bibitem [{\citenamefont {Poshakinskiy}\ and\ \citenamefont
  {Poddubny}(2019)}]{KerkerPRX}%
  \BibitemOpen
  \bibfield  {author} {\bibinfo {author} {\bibfnamefont {A.~V.}\ \bibnamefont
  {Poshakinskiy}}\ and\ \bibinfo {author} {\bibfnamefont {A.~N.}\ \bibnamefont
  {Poddubny}},\ }\bibfield  {title} {\enquote {\bibinfo {title} {Optomechanical
  kerker effect},}\ }\href {\doibase 10.1103/PhysRevX.9.011008} {\bibfield
  {journal} {\bibinfo  {journal} {Phys. Rev. X}\ }\textbf {\bibinfo {volume}
  {9}},\ \bibinfo {pages} {011008} (\bibinfo {year} {2019})}\BibitemShut
  {NoStop}%
\bibitem [{\citenamefont {Aronovitz}\ and\ \citenamefont
  {Lubensky}(1988)}]{Aronovitz1988}%
  \BibitemOpen
  \bibfield  {author} {\bibinfo {author} {\bibfnamefont {J.~A.}\ \bibnamefont
  {Aronovitz}}\ and\ \bibinfo {author} {\bibfnamefont {T.~C.}\ \bibnamefont
  {Lubensky}},\ }\bibfield  {title} {\enquote {\bibinfo {title} {Fluctuations
  of solid membranes},}\ }\href {\doibase 10.1103/PhysRevLett.60.2634}
  {\bibfield  {journal} {\bibinfo  {journal} {Phys. Rev. Lett.}\ }\textbf
  {\bibinfo {volume} {60}},\ \bibinfo {pages} {2634} (\bibinfo {year}
  {1988})}\BibitemShut {NoStop}%
\bibitem [{\citenamefont {Le~Doussal}\ and\ \citenamefont
  {Radzihovsky}(1992)}]{Doussal1992}%
  \BibitemOpen
  \bibfield  {author} {\bibinfo {author} {\bibfnamefont {P.}~\bibnamefont
  {Le~Doussal}}\ and\ \bibinfo {author} {\bibfnamefont {L.}~\bibnamefont
  {Radzihovsky}},\ }\bibfield  {title} {\enquote {\bibinfo {title}
  {Self-consistent theory of polymerized membranes},}\ }\href {\doibase
  10.1103/PhysRevLett.69.1209} {\bibfield  {journal} {\bibinfo  {journal}
  {Phys. Rev. Lett.}\ }\textbf {\bibinfo {volume} {69}},\ \bibinfo {pages}
  {1209} (\bibinfo {year} {1992})}\BibitemShut {NoStop}%
\bibitem [{\citenamefont {Ivanov}\ and\ \citenamefont
  {{Keldysh}}(1982)}]{Ivanov1982}%
  \BibitemOpen
  \bibfield  {author} {\bibinfo {author} {\bibfnamefont {A.~L.}\ \bibnamefont
  {Ivanov}}\ and\ \bibinfo {author} {\bibfnamefont {L.}~\bibnamefont
  {{Keldysh}}},\ }\bibfield  {title} {\enquote {\bibinfo {title} {Restructuring
  of polariton and phonon spectra of a semiconductor in the presence of a
  strong electromagnetic wave},}\ }\href@noop {} {\bibfield  {journal}
  {\bibinfo  {journal} {Zh. Eksp. Teor. Fiz.}\ }\textbf {\bibinfo {volume}
  {84}},\ \bibinfo {pages} {404} (\bibinfo {year} {1982})},\ \translation{Sov.
  Phys. JETP {\bf 57}, 234 (1983)}\BibitemShut {NoStop}%
\bibitem [{\citenamefont {Poshakinskiy}\ \emph {et~al.}(2016)\citenamefont
  {Poshakinskiy}, \citenamefont {Poddubny},\ and\ \citenamefont
  {Fainstein}}]{Poshakinskiy2016PRL}%
  \BibitemOpen
  \bibfield  {author} {\bibinfo {author} {\bibfnamefont {A.~V.}\ \bibnamefont
  {Poshakinskiy}}, \bibinfo {author} {\bibfnamefont {A.~N.}\ \bibnamefont
  {Poddubny}}, \ and\ \bibinfo {author} {\bibfnamefont {A.}~\bibnamefont
  {Fainstein}},\ }\bibfield  {title} {\enquote {\bibinfo {title} {Multiple
  quantum wells for $\mathcal{P}\mathcal{T}$-symmetric phononic crystals},}\
  }\href {\doibase 10.1103/PhysRevLett.117.224302} {\bibfield  {journal}
  {\bibinfo  {journal} {Phys. Rev. Lett.}\ }\textbf {\bibinfo {volume} {117}},\
  \bibinfo {pages} {224302} (\bibinfo {year} {2016})}\BibitemShut {NoStop}%
\bibitem [{\citenamefont {Poshakinskiy}\ and\ \citenamefont
  {Poddubny}(2017)}]{Poshakinskiy2017}%
  \BibitemOpen
  \bibfield  {author} {\bibinfo {author} {\bibfnamefont {A.~V.}\ \bibnamefont
  {Poshakinskiy}}\ and\ \bibinfo {author} {\bibfnamefont {A.~N.}\ \bibnamefont
  {Poddubny}},\ }\bibfield  {title} {\enquote {\bibinfo {title} {Phonoritonic
  crystals with a synthetic magnetic field for an acoustic diode},}\ }\href
  {\doibase 10.1103/PhysRevLett.118.156801} {\bibfield  {journal} {\bibinfo
  {journal} {Phys. Rev. Lett.}\ }\textbf {\bibinfo {volume} {118}},\ \bibinfo
  {pages} {156801} (\bibinfo {year} {2017})}\BibitemShut {NoStop}%
\bibitem [{Note1()}]{Note1}%
  \BibitemOpen
  \bibinfo {note} {A similar expression for the optomechanical damping rate
  correction in the system of coupled waveguide and vibrating resonant particle
  was derived in~\cite {KerkerPRX}}\BibitemShut {NoStop}%
\bibitem [{\citenamefont {Robert}\ \emph {et~al.}(2016)\citenamefont {Robert},
  \citenamefont {Lagarde}, \citenamefont {Cadiz}, \citenamefont {Wang},
  \citenamefont {Lassagne}, \citenamefont {Amand}, \citenamefont {Balocchi},
  \citenamefont {Renucci}, \citenamefont {Tongay}, \citenamefont {Urbaszek},\
  and\ \citenamefont {Marie}}]{Robert2016}%
  \BibitemOpen
  \bibfield  {author} {\bibinfo {author} {\bibfnamefont {C.}~\bibnamefont
  {Robert}}, \bibinfo {author} {\bibfnamefont {D.}~\bibnamefont {Lagarde}},
  \bibinfo {author} {\bibfnamefont {F.}~\bibnamefont {Cadiz}}, \bibinfo
  {author} {\bibfnamefont {G.}~\bibnamefont {Wang}}, \bibinfo {author}
  {\bibfnamefont {B.}~\bibnamefont {Lassagne}}, \bibinfo {author}
  {\bibfnamefont {T.}~\bibnamefont {Amand}}, \bibinfo {author} {\bibfnamefont
  {A.}~\bibnamefont {Balocchi}}, \bibinfo {author} {\bibfnamefont
  {P.}~\bibnamefont {Renucci}}, \bibinfo {author} {\bibfnamefont
  {S.}~\bibnamefont {Tongay}}, \bibinfo {author} {\bibfnamefont
  {B.}~\bibnamefont {Urbaszek}}, \ and\ \bibinfo {author} {\bibfnamefont
  {X.}~\bibnamefont {Marie}},\ }\bibfield  {title} {\enquote {\bibinfo {title}
  {Exciton radiative lifetime in transition metal dichalcogenide monolayers},}\
  }\href {\doibase 10.1103/PhysRevB.93.205423} {\bibfield  {journal} {\bibinfo
  {journal} {Phys. Rev. B}\ }\textbf {\bibinfo {volume} {93}},\ \bibinfo
  {pages} {205423} (\bibinfo {year} {2016})}\BibitemShut {NoStop}%
\bibitem [{\citenamefont {Aspelmeyer}\ \emph {et~al.}(2014)\citenamefont
  {Aspelmeyer}, \citenamefont {Kippenberg},\ and\ \citenamefont
  {Marquardt}}]{Kippenberg2014}%
  \BibitemOpen
  \bibfield  {author} {\bibinfo {author} {\bibfnamefont {M.}~\bibnamefont
  {Aspelmeyer}}, \bibinfo {author} {\bibfnamefont {T.~J.}\ \bibnamefont
  {Kippenberg}}, \ and\ \bibinfo {author} {\bibfnamefont {F.}~\bibnamefont
  {Marquardt}},\ }\bibfield  {title} {\enquote {\bibinfo {title} {Cavity
  optomechanics},}\ }\href {\doibase 10.1103/RevModPhys.86.1391} {\bibfield
  {journal} {\bibinfo  {journal} {Rev. Mod. Phys.}\ }\textbf {\bibinfo {volume}
  {86}},\ \bibinfo {pages} {1391} (\bibinfo {year} {2014})}\BibitemShut
  {NoStop}%
\bibitem [{\citenamefont {Burmistrov}\ and\ \citenamefont
  {Kachorovskii}()}]{BurKach}%
  \BibitemOpen
  \bibfield  {author} {\bibinfo {author} {\bibfnamefont {I.~S.}\ \bibnamefont
  {Burmistrov}}\ and\ \bibinfo {author} {\bibfnamefont {V.~Y.}\ \bibnamefont
  {Kachorovskii}},\ }\href@noop {} {}\bibinfo {howpublished} {private
  communication}\BibitemShut {NoStop}%
\bibitem [{\citenamefont {Roberts}\ \emph {et~al.}(2011)\citenamefont
  {Roberts}, \citenamefont {Cormode}, \citenamefont {Reynolds}, \citenamefont
  {Newhouse-Illige}, \citenamefont {LeRoy},\ and\ \citenamefont
  {Sandhu}}]{Roberts2011}%
  \BibitemOpen
  \bibfield  {author} {\bibinfo {author} {\bibfnamefont {A.}~\bibnamefont
  {Roberts}}, \bibinfo {author} {\bibfnamefont {D.}~\bibnamefont {Cormode}},
  \bibinfo {author} {\bibfnamefont {C.}~\bibnamefont {Reynolds}}, \bibinfo
  {author} {\bibfnamefont {T.}~\bibnamefont {Newhouse-Illige}}, \bibinfo
  {author} {\bibfnamefont {B.~J.}\ \bibnamefont {LeRoy}}, \ and\ \bibinfo
  {author} {\bibfnamefont {A.~S.}\ \bibnamefont {Sandhu}},\ }\bibfield  {title}
  {\enquote {\bibinfo {title} {Response of graphene to femtosecond
  high-intensity laser irradiation},}\ }\href {\doibase 10.1063/1.3623760}
  {\bibfield  {journal} {\bibinfo  {journal} {Appl. Phys. Lett.}\ }\textbf
  {\bibinfo {volume} {99}},\ \bibinfo {pages} {051912} (\bibinfo {year}
  {2011})}\BibitemShut {NoStop}%
\bibitem [{\citenamefont {Buchmann}\ \emph {et~al.}(2012)\citenamefont
  {Buchmann}, \citenamefont {Zhang}, \citenamefont {Chiruvelli},\ and\
  \citenamefont {Meystre}}]{Buchmann2012}%
  \BibitemOpen
  \bibfield  {author} {\bibinfo {author} {\bibfnamefont {L.~F.}\ \bibnamefont
  {Buchmann}}, \bibinfo {author} {\bibfnamefont {L.}~\bibnamefont {Zhang}},
  \bibinfo {author} {\bibfnamefont {A.}~\bibnamefont {Chiruvelli}}, \ and\
  \bibinfo {author} {\bibfnamefont {P.}~\bibnamefont {Meystre}},\ }\bibfield
  {title} {\enquote {\bibinfo {title} {Macroscopic tunneling of a membrane in
  an optomechanical double-well potential},}\ }\href {\doibase
  10.1103/PhysRevLett.108.210403} {\bibfield  {journal} {\bibinfo  {journal}
  {Phys. Rev. Lett.}\ }\textbf {\bibinfo {volume} {108}},\ \bibinfo {pages}
  {210403} (\bibinfo {year} {2012})}\BibitemShut {NoStop}%
\bibitem [{\citenamefont {{Xu}}\ \emph {et~al.}(2017)\citenamefont {{Xu}},
  \citenamefont {{Kemiktarak}}, \citenamefont {{Fan}}, \citenamefont
  {{Ragole}}, \citenamefont {{Lawall}},\ and\ \citenamefont
  {{Taylor}}}]{Xu2017}%
  \BibitemOpen
  \bibfield  {author} {\bibinfo {author} {\bibfnamefont {H.}~\bibnamefont
  {{Xu}}}, \bibinfo {author} {\bibfnamefont {U.}~\bibnamefont {{Kemiktarak}}},
  \bibinfo {author} {\bibfnamefont {J.}~\bibnamefont {{Fan}}}, \bibinfo
  {author} {\bibfnamefont {S.}~\bibnamefont {{Ragole}}}, \bibinfo {author}
  {\bibfnamefont {J.}~\bibnamefont {{Lawall}}}, \ and\ \bibinfo {author}
  {\bibfnamefont {J.~M.}\ \bibnamefont {{Taylor}}},\ }\bibfield  {title}
  {\enquote {\bibinfo {title} {{Observation of optomechanical buckling
  transitions}},}\ }\href {\doibase 10.1038/ncomms14481} {\bibfield  {journal}
  {\bibinfo  {journal} {Nat. Commun.}\ }\textbf {\bibinfo {volume} {8}},\
  \bibinfo {eid} {14481} (\bibinfo {year} {2017})}\BibitemShut {NoStop}%
\bibitem [{\citenamefont {Camerer}\ \emph {et~al.}(2011)\citenamefont
  {Camerer}, \citenamefont {Korppi}, \citenamefont {J\"ockel}, \citenamefont
  {Hunger}, \citenamefont {H\"ansch},\ and\ \citenamefont
  {Treutlein}}]{Camerer2011}%
  \BibitemOpen
  \bibfield  {author} {\bibinfo {author} {\bibfnamefont {S.}~\bibnamefont
  {Camerer}}, \bibinfo {author} {\bibfnamefont {M.}~\bibnamefont {Korppi}},
  \bibinfo {author} {\bibfnamefont {A.}~\bibnamefont {J\"ockel}}, \bibinfo
  {author} {\bibfnamefont {D.}~\bibnamefont {Hunger}}, \bibinfo {author}
  {\bibfnamefont {T.~W.}\ \bibnamefont {H\"ansch}}, \ and\ \bibinfo {author}
  {\bibfnamefont {P.}~\bibnamefont {Treutlein}},\ }\bibfield  {title} {\enquote
  {\bibinfo {title} {Realization of an optomechanical interface between
  ultracold atoms and a membrane},}\ }\href {\doibase
  10.1103/PhysRevLett.107.223001} {\bibfield  {journal} {\bibinfo  {journal}
  {Phys. Rev. Lett.}\ }\textbf {\bibinfo {volume} {107}},\ \bibinfo {pages}
  {223001} (\bibinfo {year} {2011})}\BibitemShut {NoStop}%
\bibitem [{\citenamefont {Vochezer}\ \emph {et~al.}(2018)\citenamefont
  {Vochezer}, \citenamefont {Kampschulte}, \citenamefont {Hammerer},\ and\
  \citenamefont {Treutlein}}]{Vochezer2018}%
  \BibitemOpen
  \bibfield  {author} {\bibinfo {author} {\bibfnamefont {A.}~\bibnamefont
  {Vochezer}}, \bibinfo {author} {\bibfnamefont {T.}~\bibnamefont
  {Kampschulte}}, \bibinfo {author} {\bibfnamefont {K.}~\bibnamefont
  {Hammerer}}, \ and\ \bibinfo {author} {\bibfnamefont {P.}~\bibnamefont
  {Treutlein}},\ }\bibfield  {title} {\enquote {\bibinfo {title}
  {Light-mediated collective atomic motion in an optical lattice coupled to a
  membrane},}\ }\href {\doibase 10.1103/PhysRevLett.120.073602} {\bibfield
  {journal} {\bibinfo  {journal} {Phys. Rev. Lett.}\ }\textbf {\bibinfo
  {volume} {120}},\ \bibinfo {pages} {073602} (\bibinfo {year}
  {2018})}\BibitemShut {NoStop}%
\bibitem [{\citenamefont {Hochmuth}\ \emph {et~al.}(1996)\citenamefont
  {Hochmuth}, \citenamefont {Shao}, \citenamefont {Dai},\ and\ \citenamefont
  {Sheetz}}]{HOCHMUTH1996}%
  \BibitemOpen
  \bibfield  {author} {\bibinfo {author} {\bibfnamefont {F.}~\bibnamefont
  {Hochmuth}}, \bibinfo {author} {\bibfnamefont {J.}~\bibnamefont {Shao}},
  \bibinfo {author} {\bibfnamefont {J.}~\bibnamefont {Dai}}, \ and\ \bibinfo
  {author} {\bibfnamefont {M.}~\bibnamefont {Sheetz}},\ }\bibfield  {title}
  {\enquote {\bibinfo {title} {Deformation and flow of membrane into tethers
  extracted from neuronal growth cones},}\ }\href {\doibase
  https://doi.org/10.1016/S0006-3495(96)79577-2} {\bibfield  {journal}
  {\bibinfo  {journal} {Biophysical Journal}\ }\textbf {\bibinfo {volume}
  {70}},\ \bibinfo {pages} {358} (\bibinfo {year} {1996})}\BibitemShut
  {NoStop}%
\bibitem [{\citenamefont {Pontes}\ \emph {et~al.}(2017)\citenamefont {Pontes},
  \citenamefont {Monzo},\ and\ \citenamefont {Gauthier}}]{Pontes2017}%
  \BibitemOpen
  \bibfield  {author} {\bibinfo {author} {\bibfnamefont {B.}~\bibnamefont
  {Pontes}}, \bibinfo {author} {\bibfnamefont {P.}~\bibnamefont {Monzo}}, \
  and\ \bibinfo {author} {\bibfnamefont {N.~C.}\ \bibnamefont {Gauthier}},\
  }\bibfield  {title} {\enquote {\bibinfo {title} {Membrane tension: A
  challenging but universal physical parameter in cell biology},}\ }\href
  {\doibase https://doi.org/10.1016/j.semcdb.2017.08.030} {\bibfield  {journal}
  {\bibinfo  {journal} {Seminars in Cell \& Developmental Biology}\ }\textbf
  {\bibinfo {volume} {71}},\ \bibinfo {pages} {30} (\bibinfo {year}
  {2017})}\BibitemShut {NoStop}%
\bibitem [{\citenamefont {Atwater}\ \emph {et~al.}(2018)\citenamefont
  {Atwater}, \citenamefont {Davoyan}, \citenamefont {Ilic}, \citenamefont
  {Jariwala}, \citenamefont {Sherrott}, \citenamefont {Went}, \citenamefont
  {Whitney},\ and\ \citenamefont {Wong}}]{Atwater2018}%
  \BibitemOpen
  \bibfield  {author} {\bibinfo {author} {\bibfnamefont {H.~A.}\ \bibnamefont
  {Atwater}}, \bibinfo {author} {\bibfnamefont {A.~R.}\ \bibnamefont
  {Davoyan}}, \bibinfo {author} {\bibfnamefont {O.}~\bibnamefont {Ilic}},
  \bibinfo {author} {\bibfnamefont {D.}~\bibnamefont {Jariwala}}, \bibinfo
  {author} {\bibfnamefont {M.~C.}\ \bibnamefont {Sherrott}}, \bibinfo {author}
  {\bibfnamefont {C.~M.}\ \bibnamefont {Went}}, \bibinfo {author}
  {\bibfnamefont {W.~S.}\ \bibnamefont {Whitney}}, \ and\ \bibinfo {author}
  {\bibfnamefont {J.}~\bibnamefont {Wong}},\ }\bibfield  {title} {\enquote
  {\bibinfo {title} {Materials challenges for the starshot lightsail},}\ }\href
  {\doibase 10.1038/s41563-018-0075-8} {\bibfield  {journal} {\bibinfo
  {journal} {Nat. Mater.}\ }\textbf {\bibinfo {volume} {17}},\ \bibinfo {pages}
  {861} (\bibinfo {year} {2018})}\BibitemShut {NoStop}%
\bibitem [{\citenamefont {Ilic}\ and\ \citenamefont
  {Atwater}(2019)}]{Ilic2019}%
  \BibitemOpen
  \bibfield  {author} {\bibinfo {author} {\bibfnamefont {O.}~\bibnamefont
  {Ilic}}\ and\ \bibinfo {author} {\bibfnamefont {H.~A.}\ \bibnamefont
  {Atwater}},\ }\bibfield  {title} {\enquote {\bibinfo {title}
  {Self-stabilizing photonic levitation and propulsion of nanostructured
  macroscopic objects},}\ }\href {\doibase 10.1038/s41566-019-0373-y}
  {\bibfield  {journal} {\bibinfo  {journal} {Nat. Photonics}\ }\textbf
  {\bibinfo {volume} {13}},\ \bibinfo {pages} {289} (\bibinfo {year}
  {2019})}\BibitemShut {NoStop}%
\bibitem [{\citenamefont {Pfeifer}\ \emph {et~al.}(2007)\citenamefont
  {Pfeifer}, \citenamefont {Nieminen}, \citenamefont {Heckenberg},\ and\
  \citenamefont {Rubinsztein-Dunlop}}]{Pfeifer07}%
  \BibitemOpen
  \bibfield  {author} {\bibinfo {author} {\bibfnamefont {R.~N.~C.}\
  \bibnamefont {Pfeifer}}, \bibinfo {author} {\bibfnamefont {T.~A.}\
  \bibnamefont {Nieminen}}, \bibinfo {author} {\bibfnamefont {N.~R.}\
  \bibnamefont {Heckenberg}}, \ and\ \bibinfo {author} {\bibfnamefont
  {H.}~\bibnamefont {Rubinsztein-Dunlop}},\ }\bibfield  {title} {\enquote
  {\bibinfo {title} {Colloquium: Momentum of an electromagnetic wave in
  dielectric media},}\ }\href {\doibase 10.1103/RevModPhys.79.1197} {\bibfield
  {journal} {\bibinfo  {journal} {Rev. Mod. Phys.}\ }\textbf {\bibinfo {volume}
  {79}},\ \bibinfo {pages} {1197} (\bibinfo {year} {2007})}\BibitemShut
  {NoStop}%
\end{thebibliography}
\end{document}